\documentclass[fleqn,usenatbib]{mnras}

\usepackage{array}
\usepackage{subfig}
\usepackage{bm}
\usepackage{newtxtext,newtxmath}
\usepackage[T1]{fontenc}
\usepackage{ae,aecompl}
\usepackage{graphicx}
\usepackage{amsmath}
\usepackage{subfig}
\usepackage{float}
\usepackage{threeparttable}
\usepackage{comment}
\restylefloat{figure}
\usepackage{url}
\usepackage{hyperref}
\usepackage[labelfont=bf,labelsep=period]{caption}
\DeclareRobustCommand{\VAN}[3]{#2}
\let\VANthebibliography\thebibliography
\def\thebibliography{\DeclareRobustCommand{\VAN}[3]{##3}\VANthebibliography}

\def\arcsec{\hbox{$^{\prime\prime}$}}
\def\approxlt{\ifmmode \rlap{$<$}{}_{{}_{{}_{\textstyle\sim}}} \else%
$\rlap{$<$}{}_{{}_{{}_{\textstyle\sim}}}$\fi}
\def\chan{{\it Chandra}}
\def\xmm{XMM-{\it Newton}}

\def\farcs{\hbox{$.\!\!^{\prime\prime}$}}

\def\arcsec{\hbox{$^{\prime\prime}$}}

\def\farcs{\hbox{$.\!\!^{\prime\prime}$}}

\usepackage{hyperref}
\usepackage{academicons}
\usepackage{xcolor}
\usepackage{orcidlink}

\title[The Rapidly Spinning IMBH J150052]{The Rapidly Spinning Intermediate--Mass Black Hole 3XMM~J150052.0+015452}
\author[Cao et al.]{Z.~Cao,\orcidlink{0000-0002-0588-6555}$^{1,2}$\thanks{E-mail: z.cao@sron.nl},
P.G.~Jonker,\orcidlink{0000-0001-5679-0695}$^{2,1}$, S.~Wen,\orcidlink{0000-0002-0934-2686}$^{2,4}$, N.C.~Stone,\orcidlink{0000-0002-4337-9458}$^3$, A.I.~Zabludoff,\orcidlink{0000-0001-6047-8469}$^4$\\
$^1$SRON, Netherlands Institute for Space Research, Niels Bohrweg 4, 2333 CA Leiden, The Netherlands\\
$^2$Department of Astrophysics/IMAPP, Radboud University, P.O.~Box 9010, 6500 GL, Nijmegen, The Netherlands\\
$^3$Racah Institute of Physics, The Hebrew University, Jerusalem, 91904, Israel\\
$^4$The University of Arizona, 933 N. Cherry Ave., Tucson, AZ  85721
}

\date{Accepted XXX. Received YYY; in original form ZZZ}

\pubyear{xxx}

\begin{document}
\label{firstpage}
\pagerange{\pageref{firstpage}--\pageref{lastpage}}
\maketitle

% Abstract of the paper
\begin{abstract}
A star tidally disrupted by a black hole can form an accretion disc with a super--Eddington mass accretion rate; the X-ray emission produced by the inner disc provides constraints on the black hole mass
$M_\bullet$ and dimensionless spin parameter $a_\bullet$.
%. The X-ray spectrum can be used to determine the black hole mass $M_\bullet$ and dimensionless spin parameter $a_\bullet$.
Previous studies have suggested that the $M_\bullet$ responsible for the tidal disruption event 3XMM~J150052.0+015452 (hereafter J150052) is $\sim$10$^{5}$~$M_{\odot}$, in the intermediate black hole (IMBH) regime. Fitting multi-epoch \xmm{} and \chan{} X-ray spectra obtained after 2008 during the source's decade-long decay, with our latest slim accretion disc model gives
$M_\bullet = 2.0^{+1.0}_{-0.3}\times10^{5}$~$M_{\odot}$ (at 
68\% confidence)
and $a_\bullet > 0.97$ (a 84.1\% confidence lower limit). The spectra obtained between 2008--2014 are significantly harder than those after 2014, an evolution that can be well explained by including the effects of inverse--Comptonisation by a corona on the early--time spectra. The corona is present when the source accretion rate is super--Eddington, while
there is 
no evidence for its effect 
in data obtained after 2014, when the mass accretion rate is around the Eddington--limit. Based on our spectral study, we infer that the corona is optically thick and warm ($kT_e=2.3^{+2.7}_{-0.8}$~keV). Our mass and spin measurements of J150052 confirm it as an IMBH and point to a rapid, near extremal, spin. These $M_\bullet$ and $a_\bullet$ values rule out both vector bosons and axions of masses $\sim10^{-16}$~eV.

\end{abstract}

% Select between one and six entries from the list of approved keywords.
% Don't make up new ones.

%%%%%%%%%%%%%%%%%%%%%%%%%%%%%%%%%%%%%%%%%%%%%%%%%%

%%%%%%%%%%%%%%%%% BODY OF PAPER %%%%%%%%%%%%%%%%%%

\section{Introduction}
\label{sec:intro}
A star approaching a black hole (BH) can be broken apart by tidal forces, leading to a tidal disruption event (TDE; e.g., \citealt{hills1975possible,rees1988tidal}). The stellar debris from the disrupted star can be subsequently accreted by the BH through an accretion disc. The electromagnetic flares associated with TDEs induced by supermassive black holes (SMBH; $\gtrsim10^{6}$~$M_{\odot}$) are mainly observed in the optical/UV and X-ray energy bands \citep[e.g.,][]{bade1996detection,komossa2004huge,gezari2006ultraviolet,van2011optical,saxton2014x,van2020optical,saxton2020x}. The TDE X-ray spectrum is often dominated by soft X-ray thermal emission \citep[][]{ulmer1999flares,lodato2011multiband}. Therefore, it has been proposed that the X-ray data of TDEs can be used to constrain the mass and the spin of their host BHs due to the high sensitivity of the disc emission to these two BH properties \citep[e.g.,][]{wen2020continuum}.

Studying TDEs provides a unique opportunity to find intermediate--mass black holes (IMBHs; $10^{2}\approxlt M_\bullet\approxlt 10^{6}$~$M_{\odot}$) and to constrain their properties, because the volumetric rate of TDEs is predicted to be dominated by IMBHs, should they exist in dense stellar environments \citep[][]{wang2004revised,stone2016rates}. IMBHs are believed to be important stepping-stones in the growth of SMBHs \citep[e.g.,][]{volonteri2010formation,banados2018800}. Thus, searching for IMBHs can help constrain the masses of SMBH seeds \citep[e.g.,][]{kormendy2013coevolution,shankar2016selection,pacucci2018glimmering}. %
It 
%has also been predicted 
is also expected
that IMBH mergers will be a prime source of gravitational radiation for the upcoming gravitational wave detector in space (\textit{Laser Interferometer Space Antenna} or LISA; e.g., \citealt{amaro2015research}).
However, it is still unclear how IMBHs form and evolve
%are formed and how they evolve
\citep[see][for a recent review]{inayoshi2020assembly}. 
%Searching for IMBHs in the local Universe can help constrain the masses of SMBH seeds \citep[e.g.,][]{kormendy2013coevolution,shankar2016selection,pacucci2018glimmering}. 
%It has also been predicted that IMBHs will be a prime source of gravitational radiation for the upcoming gravitational wave detector in space (\textit{Laser Interferometer Space Antenna} or LISA; e.g., \citealt{amaro2015research}).
%Nonetheless, 
Furthermore, direct measurements of their masses and spins
%the mass and spin of IMBHs
\citep[e.g.,][]{wen2021mass} are 
%still 
lacking (see, e.g., \citealt{greene2020intermediate} for a review on searching for IMBHs).

To constrain BH properties with TDEs, we need to model the TDE disc emission. The mass accretion rate of a TDE can vary by orders of magnitude on humanly accessible timescales, from sometimes highly super--Eddington to significantly sub--Eddington \citep[][]{evans1989tidal}. In near/super--Eddington phases, the inward advection of disc energy can no longer be neglected, and radiation pressure on the accretion flow makes fluid orbits non-Keplerian \citep[][]{abramowicz1988slim}. As a result, a standard ``thin'' disc model \citep[e.g.,][]{shakura1973black} is not adequate to describe such a TDE disc, and a ``slim'' disc model has to be used. Details of the slim disc solution can be found in \citet{abramowicz1988slim} and \citet{skadowski2009slim}. We have now developed such models for application to TDEs, including those associated with IMBHs \citep[][]{wen2020continuum,wen2021mass,wen2022library}.

\begin{table*}
\centering
\caption[xxx]{\xmm{} and \chan{} observations of J150052 analysed in this work. The exposure time is the time remaining after filtering for epochs of enhanced background count rates. The average count rates of the source$+$background spectra are given in the energy ranges 0.3--10.0~keV (\xmm) and 0.3--7.0~keV (\chan). We also list in the last column the source counts estimated by subtracting the estimated number of background counts in the source extraction region. We treat observations C2--C8 as a single-epoch observation,
and its estimated total source count is 6905.}
\begin{tabular}{cccccc}
\hline
Satellite  & ObsID (Label)    & Date   & Exposure (ks) & Count rate (cts/s)  &  Est.~Source counts (cts) \\
\hline
XMM-Newton & 0554680201 (X1) & 2009-02-11 & 39 & $(4.1\pm0.1)\times10^{-2}$&1305\\
           & 0554680301 (X2) & 2009-02-17 & 35 & $(4.0\pm0.1)\times10^{-2}$&1146\\
           & 0804370301 (X3) & 2017-07-21 & 14 & $(3.5\pm0.2)\times10^{-2}$&417\\
           & 0804370401 (X4) & 2017-08-09 & 5.4 & $(3.4\pm0.3)\times10^{-2}$&146\\
           & 0804370501 (X5) & 2018-01-20 & 4.5 & $(3.7\pm0.3)\times10^{-2}$&145\\
           & 0844040101 (X6) & 2020-02-21 & 20 & $(2.8\pm0.1)\times10^{-2}$&467\\
\hline
Chandra    & 9517 (C1)       & 2008-06-05 & 99 & $(1.37\pm0.04)\times10^{-2}$&1223\\
           & 12951 (C2)      & 2011-03-28 & 74 & $(1.47\pm0.06)\times10^{-2}$\\
           & 13246 (C3)      & 2011-03-30 & 45 & $(1.47\pm0.08)\times10^{-2}$\\
           & 13247 (C4)      & 2011-03-31 & 36 & $(1.62\pm0.09)\times10^{-2}$\\
           & 12952 (C5)      & 2011-04-05 & 143 & $(1.52\pm0.04)\times10^{-2}$&C2--C8: 6905\\
           & 12953 (C6)      & 2011-04-07 & 32 & $(1.67\pm0.09)\times10^{-2}$\\
           & 13253 (C7)      & 2011-04-08 & 118 & $(1.49\pm0.05)\times10^{-2}$\\
           & 13255 (C8)      & 2011-04-10 & 43 & $(1.50\pm0.08)\times10^{-2}$\\
           & 17019 (C9)      & 2015-02-23 & 37 & $(0.51\pm0.04)\times10^{-2}$&185\\
\hline
\end{tabular}
\label{tb:obs}
\end{table*}

Modelling TDE disc emission not only constrains the BH mass, but also the spin. In fact, TDE modelling is currently the only way to probe the spins of IMBHs \citep[][]{wen2020continuum,wen2021mass}.
%, providing a unique opportunity to probe the formation and evolution theory of the IMBH population (e.g., \citealt{berti2008cosmological}).
The BH spin distribution reveals how they have grown (e.g., \citealt{berti2008cosmological}), and, for individual IMBHs, how they may have formed 
%potentially how they were formed 
\citep[e.g.,][]{inayoshi2020assembly}. 
%TDE modelling is currently the only way to probe the spins of IMBHs \citep[][]{wen2020continuum,wen2021mass}, and thus it provides a unique opportunity to probe the formation and evolution theory of the IMBH population.
%Potential (systematic) uncertainties in the TDE spin measurement do exist, however: for example, if the stellar debris is not efficiently circularised \citep[e.g.,][]{hayasaki2016circularization} or if the disc is not aligned with the equatorial plane of the BH spin \citep[e.g.,][]{stone2012observing,franchini2016lense}.  
%Despite these concerns, the potential of TDEs to survey massive BH demographics is immense. 
In the near future, time domain surveys 
from the
%such as 
VRO \citep{bricman2019lsst}, {\it eROSITA} \citep{khabibullin2014erosita, jonker2020implications}, {\it Einstein Probe} \citep{2018SPIE10699E..25Y}, and {\it ULTRASAT} \citep{sagiv2014ultra} will together find thousands of TDEs. The resulting large TDE samples, when their follow--up X--ray observations are analysed through slim disc modelling, will produce constraints on the distributions of IMBH mass and spin for the first time, providing a unique opportunity to probe the formation and evolution theory of the IMBH population.

By constraining both the mass and the spin of IMBHs, we can also test for the existence of ultralight bosons, such as axions (or axion-like particles) and vector bosons. Scalar axions have long been considered a possible solution to the strong CP problem \citep{peccei1977qcd}.
%, and more 
More recently, axion-like particles have received attention as a natural consequence of string theory \citep{arvanitaki2010axiverse}. Both scalar and vector ultralight bosons are of astrophysical interest as particle dark matter candidates \citep{dine1983axion, preskill1983axion, nelson2011vectors, arias2012dark, graham2016vectors}. 
%In some cases, their condensates may form ``boson stars'' of high density that could mimic BHs \citep{liebling2017bosons}. 
A rapidly--spinning BH can interact with ultralight bosons, causing a superradiant scattering instability 
%to develop 
that spins down the BH (\citealt{bardeen1972rotating,press1972floating}; see also \citealt{brito2020superradiance} for a recent review). Efficient spindown 
%of the BH 
only occurs when bosons %with mass $m$ exist such that %$\frac{GM_{\bullet}m}{c\hbar}\sim1$ (here $\hbar$ is the reduced Planck constant); for $\frac{GM_{\bullet}m}{c\hbar}$ values away from 1, 
of the appropriate mass exist, such that their Compton wavelength is close to the size of the event horizon. Without a close match, the timescale for establishing the instability grows exponentially. Therefore, we can use the mass and spin measurements of a rapidly--spinning BH to rule out roughly one order--of--magnitude in ultralight particle mass  \citep[e.g.,][]{cardoso2018constraining}.

Furthermore, TDEs are good laboratories for studying accretion theories in the super--Eddington regime.
The X--ray spectrum of sources accreting in the so-called ultra-luminous state can sometimes be well-described by two black-bodies \citep[e.g.,][]{pinto2017ultraluminous,pinto2021xmm}, similar to some TDEs \citep[e.g.,][]{kara2018ultrafast}.
For example,
it has been proposed that ultra--luminous X--ray sources (ULXs) host stellar--mass compact objects (neutron stars or BHs) accreting at super--Eddington rates (e.g., \citealt[][]{king2001ultraluminous,roberts2007x,feng2009spectral,gladstone2009ultraluminous,kaaret2017ultraluminous}). Perhaps these ULXs are BH X--ray binaries (XRBs) accreting in an
%so-called 
ultra--luminous state \citep[e.g.,][]{gladstone2009ultraluminous,motta2012discovery,sutton2013ultraluminous}. 
%The X--ray spectrum of sources accreting in the ultra-luminous state can sometimes be well-described by two black-bodies \citep[e.g.,][]{pinto2017ultraluminous,pinto2021xmm}, similar to some TDEs \citep[e.g.,][]{kara2018ultrafast}. 
%By studying the TDE emission, we can investigate if common physical circumstances such as a super-Eddington slim disc can help explain the potential similarity in spectral shape and evolution between TDEs and other super-Eddington accreting sources such as ULXs. 
Modelling TDE X-ray emission tests whether similarities in spectral shape and evolution between TDEs and other super-Eddington accretors like ULXs could arise from common physical circumstances, e.g., a super-Eddington slim disc.

%The majority of the X--ray emission of a 
%BH--TDE
%Most X-rays produced in TDEs are thought to come from  thermal emission from the accretion disc. At certain phases during the decay of its luminosity, a TDE can also have non--thermal emission from the BH ``corona,'' where the disc photons are Compton up--scattered by hot coronal electrons. It has been proposed that the coronal temperature has a physical limit caused by the production of electron--positron pairs from high--energy photon collisions \citep[][]{svensson1982electron,guilbert1983spectral}. Further energy input will effectively increase the number of pairs instead of the temperature when approaching the pair--production limit. 
%%Evidence shows that 
%Many coronae of SMBHs in active galactic nuclei (AGN) appear to have temperatures close to the pair--production limit; thus, they could be pair--dominated \citep[][]{fabian2015properties,fabian2017properties}. Modelling the TDE emission makes it possible to examine whether a pair--dominated corona can also arise from an accreting IMBH.

The X-ray source 3XMM~J150052.0+015452 (J150052) was detected by \xmm{} and \chan{} in observations of the foreground galaxy group NGC~5813 (redshift $z$=0.0064; \citealt[][]{paturel2002comparison}) in 2005. It has a well--constrained X-ray position coincident with the center of the galaxy SDSS~J150052.07+015453.8 (redshift $z$=0.14542; \citealt[][]{lin2017likely}). Follow--up \xmm{} and \chan{} observations started in 2008, and together 
the observations
span more than a decade. All the observational evidence presented in \citet{lin2017likely,lin2022follow} suggests that J150052 is a slowly-decaying TDE. This decade--long decay distinguishes J150052 from many other faster--decaying TDEs \citep[e.g.,][]{van2021seventeen}. The long duration can be attributed to slow circularisation of the fall back material, as well as a long super--Eddington phase for a less--massive BH ($<10^{6}$~$M_{\odot}$; \citealt[][]{lin2022follow}). Both the BH--bulge scaling relation \citep[][]{graham2013m} and spectral analysis using a model for a thin accretion disc$+$corona \citep{lin2022follow} determine the mass of the BH J150052 to be 
%of order
$\sim10^{5}$~$M_{\odot}$.

Here we analyze the X-ray spectra and light curve obtained over J150052's
decade--long decay, %of J150052
considering the slim disc model for the accretion disc \citep[][]{wen2020continuum,wen2021mass}. We include all the archived \xmm{} and \chan{} X-ray data obtained since 2008. In Section 2, we describe the selected data and our data reduction. In Section 3, we present the results from our model fits, including the constraints on black hole mass and spin. In Section 4 we discuss the implications of our results for IMBH formation scenarios, the origin of the Comptonisation component, the evolution of that component, and the mass of ultralight bosons. In Section 5, we end with our conclusions. 

\section{Data and data reduction}
\label{sc:data}

We use \xmm{} and \chan{} observations of J150052 in this work. Some basic properties of those observations are listed in Table~\ref{tb:obs}. Note that we do not include observations obtained before 2006, as at those epochs the mass accretion rate in J150052 was too low for our slim disc model to apply. The labels indicating the observations differ therefore from those used in \cite{lin2017likely}. 

For the \xmm{} data reduction, we use HEASOFT (version 6.28) and SAS (version 18.0.0) with the calibration files renewed on January 5th, 2021 (CCF release: XMM-CCF-REL-380). We use the SAS command \texttt{epproc} to process the Science 0 data from \xmm/EPIC-pn. We employ the standard filtering criteria\footnote{https://www.cosmos.esa.int/web/xmm-newton/sas-thread-epic-filterbackground} for EPIC-pn data, where we require that the 10--12 keV detection rate of pattern 0 events is $<$ 0.4 counts s$^{-1}$. This way the data are cleared from periods with an enhanced background count rate. We use a circular source region of 20\arcsec{} radius centred on the source for the spectral counts extraction. The background count spectra are extracted from apertures close to the source on the same EPIC-pn detector and free from other bright sources. We use a rectangular region of 134\arcsec $\times$ 45\arcsec{} 
to extract the background in observations X1 and X2 due to the source location being close to the edge of the EPIC-pn detector, while a circular region of 50\arcsec{} radius is used for the background extraction in X3--X6. We check for the presence of photon pile-up using the SAS command \texttt{epatplot} and conclude the pile-up is not important in any of our \xmm{} observations.

During some of the \xmm~observations, one of the two MOS detectors was turned off. Therefore, for consistency,
we do not use the MOS data. We also do not use the RGS data, because the signal--to--noise ratio in the RGS detectors is too low.

For the \chan{} data reduction, we use CIAO (version 4.12). We employ the CIAO commands \texttt{chandra\_repro} and \texttt{specextract} for \chan/ACIS data filtering and spectral extraction, respectively. J150052 has a large off-axis angle for the first eight \chan~observations. Following
\citet{lin2017likely} we take the deterioration of the point spread function with off-axis angle into account: we set the radius of the circle used for the source extraction region to 16\farcs7 for C1, 13\farcs4 for C2--C8, and
1\farcs6 for C9 (during the observation labelled C9 the source is observed on-axis). The background spectra are extracted from apertures close to the source, on the same \chan/ACIS chip, and free from other bright sources. We use rectangular apertures with length $>$100\arcsec{} and width $\sim$~70\arcsec{} as the background regions in C1--C8, and a circular region of 50\arcsec{} radius in C9. Because observations C2 to C8 are obtained close in time and the source spectra did not change significantly on such short time scales, we combine the spectra from C2 to C8 using the CIAO command \texttt{combine\_spectra}. We subsequently treat the C2--C8 observations as a single epoch observation,
similar to the approach of \citet{lin2017likely}.

In this paper, we focus on the energy bands 0.3--10~keV for \xmm/EPIC-pn and 0.3--7.0~keV for \chan/ACIS. We require each source$+$background and the background spectral energy bins to have a minimum of one photon per bin. For each epoch, we first fit the background spectrum with a phenomenological model. When we fit the source$+$background spectrum, we add the best-fit background model to the fit-function describing the source$+$background, fixing the background model parameters to their best-fit values determined from the fit to the background-only spectra. The best-fit background model varies from epoch to epoch and between instruments; it consists of between 1--3 power-laws and 3--6 Gaussian components (with a full--width half--maximum or FWHM of $\sigma=0.001$~keV, less than the spectral resolution in both \xmm/EPIC-pn and \chan/ACIS instruments) that accounts for the background continuum and background fluorescence lines \citep[e.g.,][]{markevitch2003chandra,katayama2004properties}.

Throughout this paper, we carry out the spectral analyses using the \texttt{XSPEC} package (\citealt{arnaud1996xspec}; version 12.11.1), applying Poisson statistics (\citealt{cash1979parameter}; \texttt{C-STAT} in \texttt{XSPEC}). Unless otherwise specified, we quote all the parameter errors at the 1$\sigma$ (68\%) confidence level, assuming $\Delta$C-stat=1.0 and $\Delta$C-stat=2.3 for single-- and two--parameter error estimations \citep{wen2021mass}, respectively. All the spectra we present here and in the Appendix are re-binned for plotting purposes only. All models in this paper include Galactic absorption of column density $N_{H,G}=4.4\times10^{20}{\rm cm}^{-2}$ \citep{kalberla2005leiden} using the model \texttt{TBabs} \citep{wilms2000absorption}. We also consider the absorption intrinsic to the X-ray source and its host galaxy at redshift $z$=0.14542 (using the model \texttt{zTBabs}), leaving the \texttt{zTBabs} column density $N_{H,i}$ to be a free parameter. With the \texttt{energies} command in \texttt{XSPEC}, we take a logarithmic energy array of 1000 steps from 0.1 to 1000.0~keV for model calculations in place of response energy arrays, to correctly calculate the Comptonisation model when needed (see Section \ref{sc:res-compton}). Residuals of each of our joint fits are shown in Figures in the Appendix. 

\section{Results}
\label{sc:results}

\subsection{Modelling using simple phenomenological models}

\begin{figure*}
    \centering
    \subfloat[\label{fig:2bb}]{\includegraphics[width=0.5\textwidth]{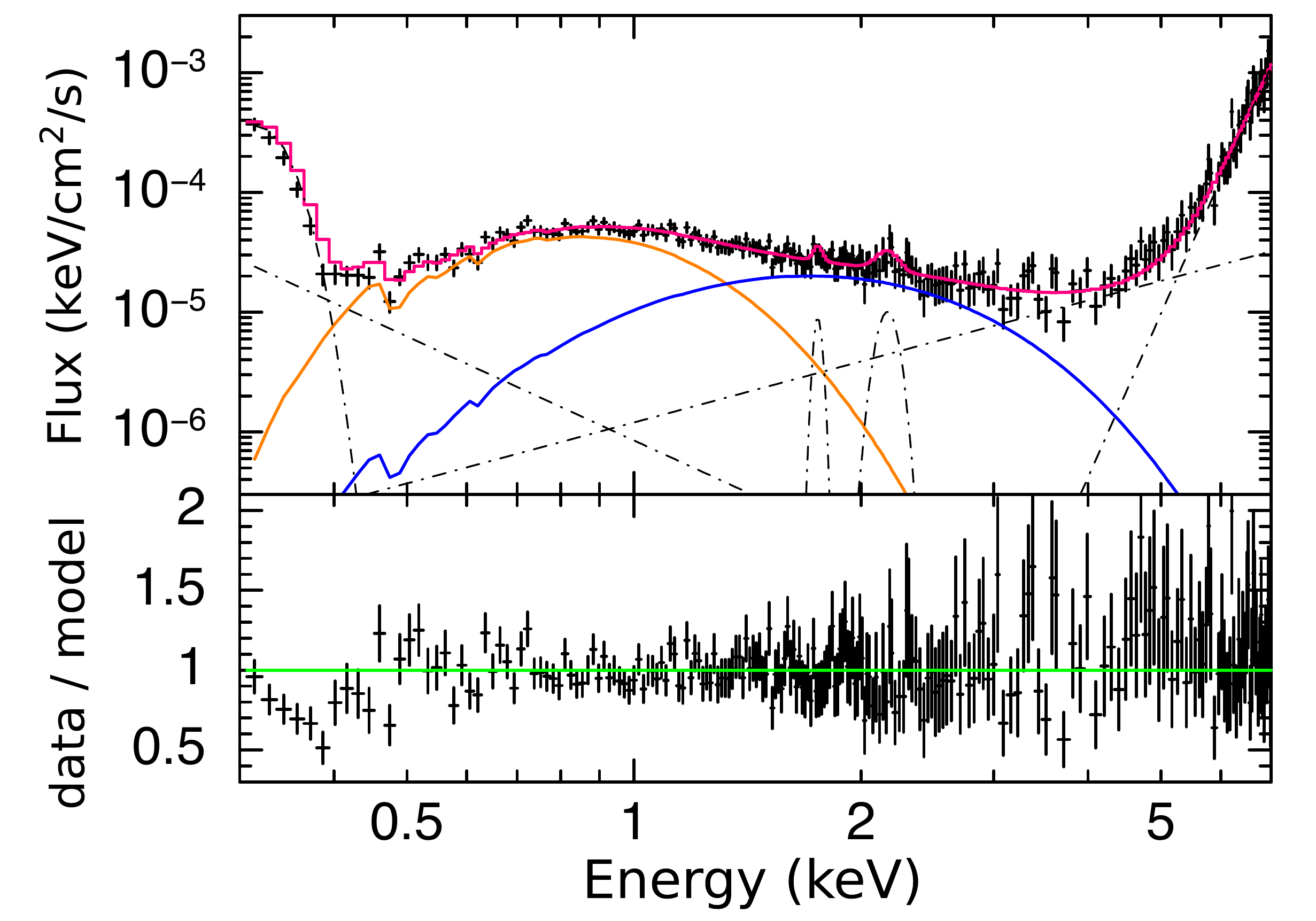}}\hfill
    \subfloat[\label{fig:1bb}]{\includegraphics[width=0.5\textwidth]{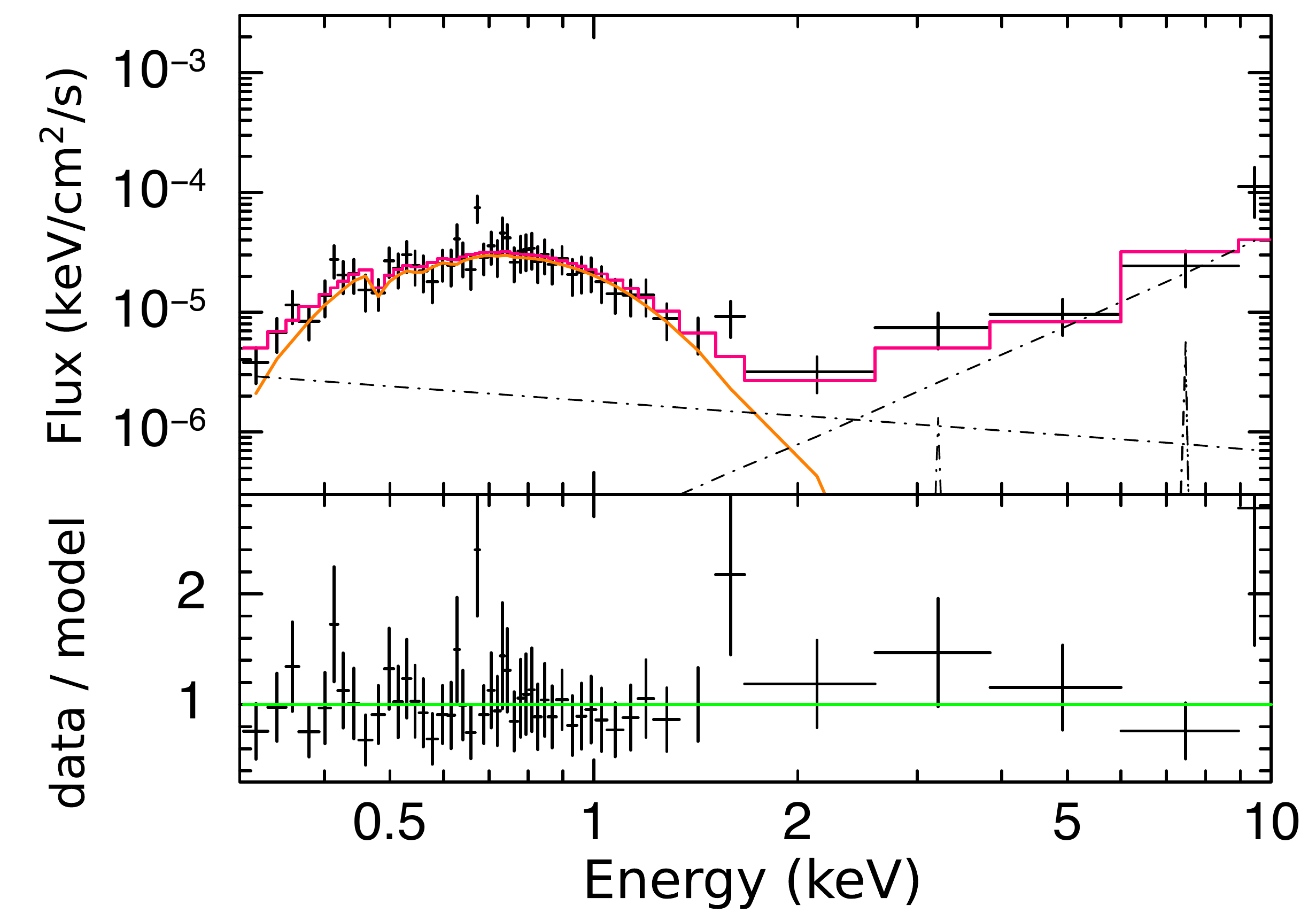}}\hfill
    \caption{\textbf{a}) The source$+$background spectrum from combining the C2--C8 observations (as one epoch), de--convolved from the detector response curve (i.e., the unfolded spectrum), and the data$/$model ratio. Here the source fit function consists of two BBs. The red line is the total source$+$background continuum, the orange and the blue lines stand for the two best-fitting BB models, and the dot--dash lines are the background model components. The parameters of the background model have been kept fixed during the fits (see Section \ref{sc:data}); \textbf{b}) Unfolded source$+$background spectrum observed at Epoch X3 and the data$/$model ratio for a source model fit function comprised of one BB. The format follows that of  Fig.~\ref{fig:2bb}, except that here only one BB model in orange is present. J150052 shows two--BB--like, hardened spectra at early epochs (C1, X1, X2, C2--C8), compared to spectra at late epochs (C9, X3, X4, X5, X6).}
    \label{fig:bb}
\end{figure*}

\begin{figure*}
    \centering
    \includegraphics[width=0.8\textwidth]{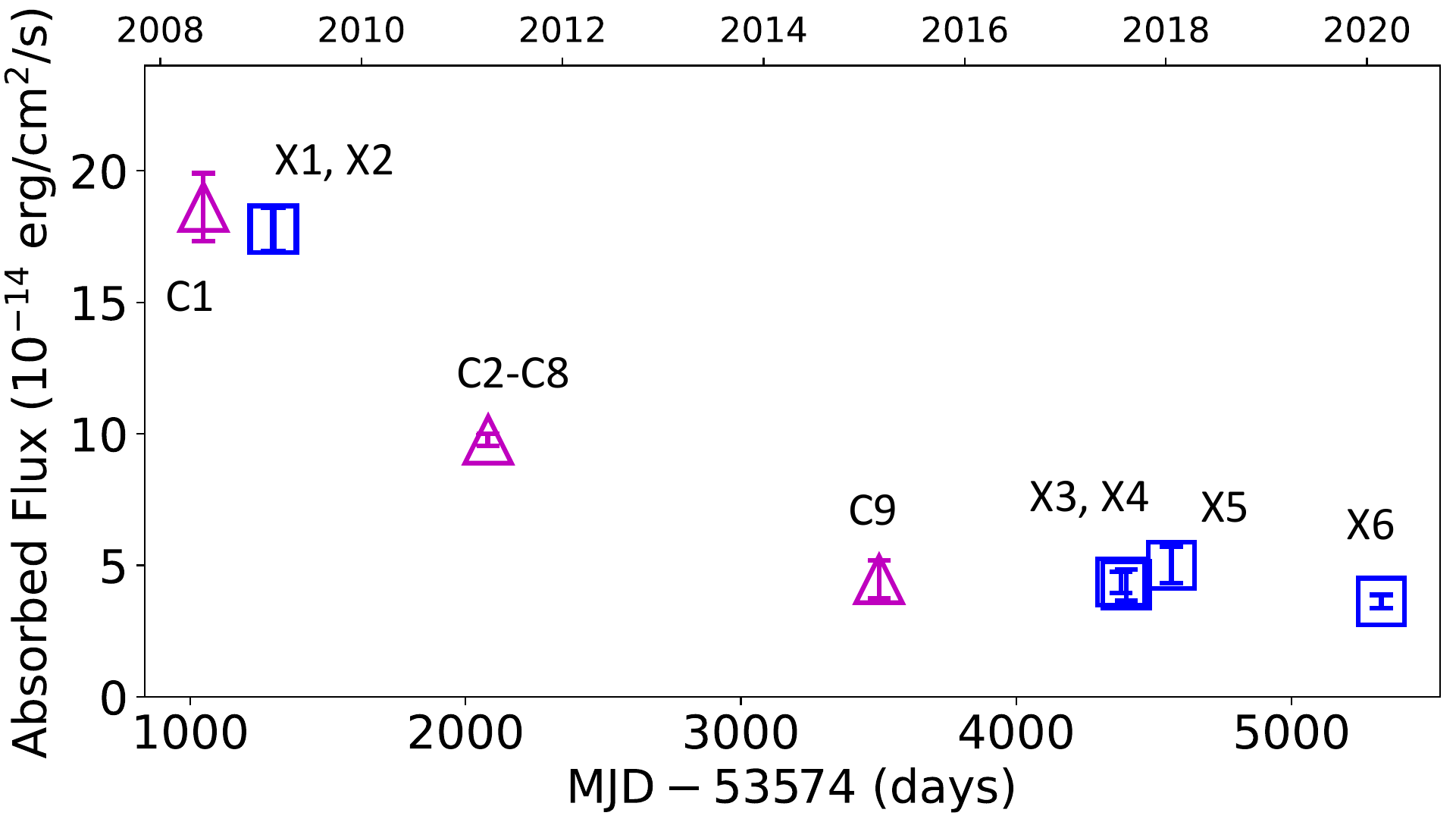}
    \caption{Long--term lightcurve of J150052 starting from 2008. The x-axis is in days since J150052's first detection on modified Julian date (MJD) 53574, with the corresponding calendar year denoted at the top of the figure. The y-axis denotes the observed flux (0.3--10~keV), including both the effects of intrinsic and Galactic absorption calculated using our two--BB fit. Blue squares are used for \xmm{} observations and purple triangles for \chan{} observations. J150052 has experienced a decade--long decay since its first detection on MJD 53574.}
    \label{fig:lcbb}
\end{figure*}

We first use phenomenological models to fit the individual spectra. Here the goal is to describe the data with a few parameters, as well as to capture any changes in the parameter values between epochs. First, we use a black body (BB) to model the spectra (\texttt{zbbody} with redshift $z$=0.14542). We find a best--fit with a total C-stat/d.o.f (degrees--of--freedom) $=2269/2162$, fitting nine epochs together while letting all parameters vary freely. However, visual inspection of the fit and the residuals shows that the best-fit single BB model does not describe the spectra well around 2~keV in several epochs (Fig.~\ref{fig:1bbr}). 

Instead, a fit function comprised of two BBs describes the data well (C-stat/d.o.f.~$=2003/2154$, Fig.~\ref{fig:2bbr}). The best-fit parameter values are given in Table~\ref{tb:2bb}. We then use the Akaike information criteria (AIC; \citealt{akaike1974new}) to investigate the significance of adding a second BB model. From the one--BB model to the two--BB model,  $\Delta$AIC $=250$ (with $\Delta$AIC$>$5 and $>$10 considered a strong and very strong improvement, respectively, over the simpler model). Therefore, we conclude that adding a second BB improves the goodness of the fits significantly. We find that the source spectra of X1, X2, C1, and C2--C8, which we refer to as the ``early epochs,'' can be well-described by two BBs, with an average temperature of $0.19\pm0.01$ keV and $0.48\pm0.03$ keV (e.g., Fig.~\ref{fig:2bb}). On the other hand, the spectra obtained at epochs C9, X3, X4, X5, and X6, a.k.a.~the ``late epochs,'' are consistent with a single BB model with a lower average temperature of $0.15\pm0.01$ keV (e.g., Fig.~\ref{fig:1bb}). 

We find that $N_{H,i}$ and the BB normalisations, A$_{\rm zbbody_{1}}$ and A$_{\rm zbbody_{2}}$, are consistent with being constant within their 3$\sigma$ uncertainties: the best-fitted overall values are $0.14\times10^{22}$cm$^{-2}$, $3.83\times10^{-6}$[$10^{37}(1+z)^{-2}$erg$/$s$/$kpc$^2$], and $7.2\times10^{-7}$[$10^{37}(1+z)^{-2}$erg$/$s$/$kpc$^2$], respectively. Fig.~\ref{fig:lcbb} shows the observed flux of J150052 as a function of time. The observed flux values are attenuated by the effect of the intrinsic and the Galactic absorption based on our two--BB fit. 

\begin{table*}
\renewcommand{\arraystretch}{1.5}
\centering
\caption[]{Best-fit parameters using the fit function \texttt{TBabs*zTBabs*(zbbody+zbbody)} to describe the source spectra. From top to bottom, epochs are listed in time sequence. The second \texttt{zbbody} model component is not necessary to obtain a good fit in C9, X3, X4, X5, X6. Therefore, it is omitted from the fit-function at these epochs.}
\begin{tabular}{cc|ccccc|c}
\hline
\multicolumn{2}{c|}{Model Component}&\texttt{zTBabs}&\multicolumn{2}{c}{\texttt{zbbody}$_1$}&\multicolumn{2}{c|}{\texttt{zbbody}$_2$}&C-stat/d.o.f\\
\multicolumn{2}{c|}{Free parameter}&$N_{H,i}$&$kT$&A$_{\rm zbbody_{1}}$&$kT$&A$_{\rm zbbody_{2}}$\\
\multicolumn{2}{c|}{Unit}&$10^{22}$~${\rm cm}^{-2}$&keV&$10^{37}(1+z)^{-2}$erg$/$s$/$kpc$^2$&keV&$10^{37}(1+z)^{-2}$erg$/$s$/$kpc$^2$\\
\hline
\hline
Early Epoch&C1&$<0.16$&$0.21^{+0.03}_{-0.02}$&$5^{+2}_{-1}\times10^{-6}$&$0.47^{+0.09}_{-0.06}$&$8^{+4}_{-3}\times10^{-7}$&179/205\\
&X1&$0.17^{+0.08}_{-0.06}$&$0.19\pm0.03$&$7^{+4}_{-2}\times10^{-6}$&$0.41^{+0.14}_{-0.09}$&$10^{+10}_{-5}\times10^{-7}$&335/396\\
&X2&$0.18^{+0.06}_{-0.05}$&$0.18\pm0.02$&$8^{+3}_{-2}\times10^{-6}$&$0.56^{+0.11}_{-0.09}$&$7\pm2\times10^{-7}$&401/392\\
&C2--C8&$0.14\pm0.04$&$0.19\pm0.01$&$30^{+7}_{-5}\times10^{-7}$&$0.48\pm0.03$&$7\pm1\times10^{-7}$&384/417\\
\hline
Late Epoch&C9&$0.14^{+0.09}_{-0.08}$&$0.14^{+0.02}_{-0.01}$&$3^{+3}_{-1}\times10^{-6}$&-&-&85/63\\
&X3&$0.11^{+0.07}_{-0.06}$&$0.16^{+0.02}_{-0.01}$&$23^{+14}_{-7}\times10^{-7}$&-&-&193/223\\
&X4&$0.3^{+0.2}_{-0.1}$&$0.12\pm0.02$&$8^{+29}_{-5}\times10^{-6}$&-&-&80/121\\
&X5&$<0.19$&$0.17^{+0.02}_{-0.03}$&$19^{+22}_{-6}\times10^{-7}$&-&-&106/115\\
&X6&$0.15\pm0.07$&$0.14\pm0.01$&$3^{+2}_{-1}\times10^{-6}$&-&-&240/222\\
\hline
\end{tabular}
\label{tb:2bb}
\end{table*}

\subsection{Slim disc modelling}
\label{sc:model}

Based on our spectral analyses using simple phenomenological models, we conclude that the shape of the continuum is changing throughout the decay of J150052 (from a two-BB shape to a single-BB shape). This behaviour might potentially be explained by the spectra at the early epochs being affected by additional spectral hardening from electron scattering and a temperature gradient in the disc atmosphere \citep[][]{shimura1993vertical,shimura1995spectral}. To constrain the TDE accretion disc parameters, as well as the mass and the spin of the black hole, we use the slim disc model %\texttt{slimfit} 
(\citealt{wen2020continuum}, updated by \citealt{wen2021mass}) to simultaneously fit the spectra at all epochs.

The slim disk %\texttt{slimfit} 
model considers the stationary, relativistic ``slim disc'' accretion disc solutions and ray-traces the disc photons self-consistently to the observer's frame. The free parameters for the slim disc are the BH mass $M_{\bullet}$, the BH dimensionless spin $a_{\bullet}$, the disc accretion rate $\dot m$, the inclination $\theta$, and the spectral hardening factor $f_c$ \citep{shimura1993vertical,shimura1995spectral}, which parameterises the spectral hardening due to electron scattering and temperature gradient in the disc atmosphere. The disc accretion rate $\dot m$ is in units of the Eddington--limited accretion rate $\dot m_{\rm Edd}$. In the model we define $\dot m_{\rm Edd}=1.37\times10^{21}$~kg~s$^{-1}\times$~$M_{\bullet}/10^{6}M_{\odot}$ \citep[][]{wen2020continuum}. Note that the actual $\dot M$ in kg~s$^{-1}$ units is not identical to the $\dot m$, and needs to be calculated after the BH mass is constrained. The model implements the astrophysical spin limit of a Kerr BH $a_{\bullet}<$0.998 \citep[][]{thorne1974disk}. The $f_c$ is expected to be $>$2, but saturates at $\sim2.4$, for near/super-Eddington accretion discs \citep{davis2006testing,davis2019spectral}. During the (simultaneous) fitting of the X-ray spectra, we let $\dot m$ 
vary between epochs, while the value of $M_{\bullet}$, $a_{\bullet}$, and $\theta$ are free to vary but are required to have the same value at each epoch. Similarly $f_c$ is treated as a single free parameter for all early epochs, while we keep it fixed to 2.2 for the late epochs. We make this latter choice 
because the late spectra 
are softer and have fewer counts,
preventing us from constraining the spectral hardening effects that mainly impact the hard spectral tail ($>1.0$~keV). As for $\{M_{\bullet}$, $a_{\bullet}\}$, we follow the approach of \citet{wen2021mass} and search the $\{M_{\bullet}$, $a_{\bullet}\}$ parameter space by performing a joint fit and minimizing the C-stat at each $\{M_{\bullet}$, $a_{\bullet}\}$ grid point. As a result, we fix $\{M_{\bullet}$, $a_{\bullet}\}$ to different values during each of the joint fits.

By jointly fitting the J150052 spectra at early and late 
times with the slim disc model, we obtain a minimum C-stat/d.o.f~$=2183/2169$ with $M_{\bullet}=1.5\times 10^5$~M$_\odot$ and $a_{\bullet}=0.998$ (Fig.~\ref{fig:fcall}). However, 
the slim disc model does not fit
the early epochs above 2~keV well, as was also the case for
the phenomenological models. Furthermore, the physical tension in this joint-fit is that $f_c$ becomes larger than 4.0 during all the early epochs, instead of saturating around 2.4
as expected in the super-Eddington regime \citep{davis2006testing}. We then test the slim disc model with $f_c$ fixed to 2.4 during the early epochs while keeping it fixed to 2.2 at late times.
We find a new minimum C-stat/d.o.f~$=3980/2170$ at $M_{\bullet}=0.5\times 10^5$~M$_\odot$ and $a_{\bullet}=0.998$. This fit is worse than the previous fit because now more hard photons above 2~keV are left un--fitted. Fig.~\ref{fig:fc24} shows the residual of this joint-fit. In both joint fits, 
the slim disc %\texttt{slimfit} 
model describes the late epochs well but it has difficulties in describing the hard spectrum observed at early epochs, especially around and above 2~keV. This discrepancy is more prominent when we fixed $f_c$ value to 2.4, as when $f_c$ is allowed to float freely the slim disc %\texttt{slimfit}
fit tries to describe the high energy photons in the spectra by increasing the disc spectral hardening, 
thereby increasing the value of $f_c$  \citep{shimura1995spectral}.

\subsection{Slim disc$+$thermal Comptonisation}
\label{sc:res-compton}

From our joint-fits with the slim disc model, we find that the early epochs of the J150052 spectra tend to be harder than a typical slim disc spectrum, which can not be solely explained by the disc spectral hardening factor $f_c$. Interestingly, this deviation from a slim disc spectrum becomes negligible at later epochs. We explore here if inverse--Comptonisation of the soft (slim disc) photons can help explain the spectral data above 2 keV. 

As a starting point, we investigate the effect of Comptonization on the slim disc photons assuming a thermal distribution of energetic electrons. These electrons can originate in a disc wind, or if present, the base of a jet. We use the convolution model \texttt{thcomp} \citep{zdziarski2020spectral} to self-consistently determine the up--scattered spectra from thermally-distributed electrons. The \texttt{thcomp} model parameterises the up--scattered spectra through the Thomson optical depth $\tau$ and the electron temperature $kT_e$ parameters. 

We fit the spectra of all epochs together and the inverse--Comptonisation component to the fit-function is only used to describe the spectra of the early epochs, as the late-epoch spectra can be well-described using a fit-function comprised of only the slim disc model (Fig.~\ref{fig:fc24}). The spectral hardening $f_c$ in slim disc %\texttt{slimfit}
is free--to--vary between 1.0 and 2.4 during the early epochs. Following section \ref{sc:model}, we fixed the $f_c$ to a value of 2.2 at the late epochs. Based on the stand--alone, depreciated, Comptonisation \texttt{nthComp} model for thin disc accretion \citep[][]{zdziarski1996broad,zycki19991989}, \citet{lin2022follow} infer the optical depth $\tau$ in the corona could be varying between early epochs. Therefore, here for the early epochs we fix the $kT_e$ to be the same but let $\tau$ be free--to--vary between the early epochs. Finally, we fix the covering fraction of \texttt{thcomp} to unity during the early epochs, so that all seed photons are going through the Comptonising cloud.

\begin{figure}
    \centering
    \includegraphics[width=\linewidth]{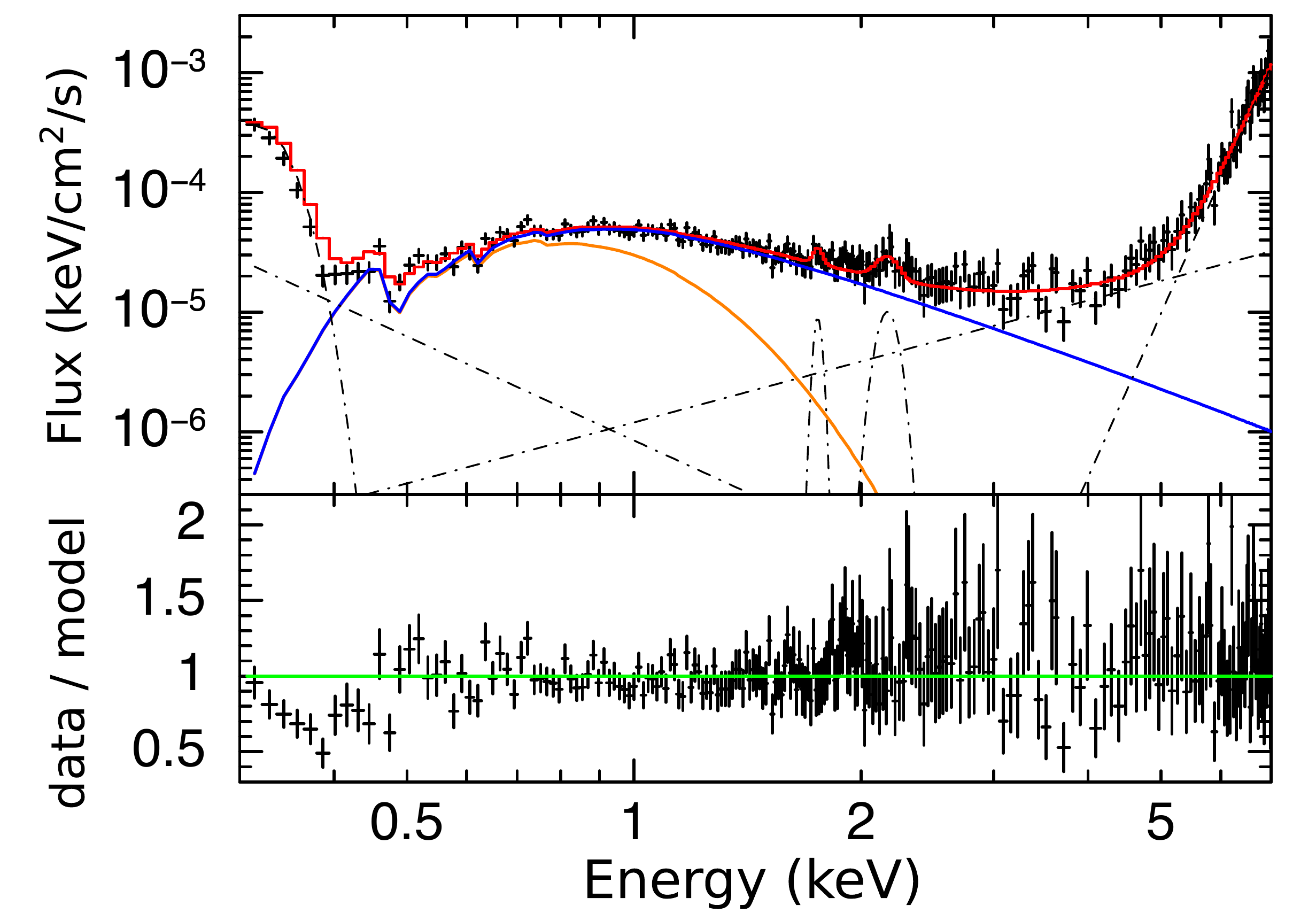}
    \caption{Unfolded source$+$background spectrum observed at Epoch C2--C8 with data$/$model residuals (the data points correspond to those shown in magenta in Fig.~\ref{fig:res} but here they are shown with a slightly different re-binning). The observed source$+$background spectrum is shown together with the best-fit slim disc model convolved with a thermal electron Comptonisation model. The format follows Fig.~\ref{fig:2bb}, except that here the orange line shows the slim disc continuum before Comptonisation and the blue line shows the total source continuum spectrum with the effect of scenario--dependent Comptonisation included. The slim disc$+$thermal Comptonisation model can describe the observed early epoch spectra of J150052 well.}
    \label{fig:uf_c}
\end{figure}

\begin{figure}
    \centering
    \includegraphics[width=\linewidth]{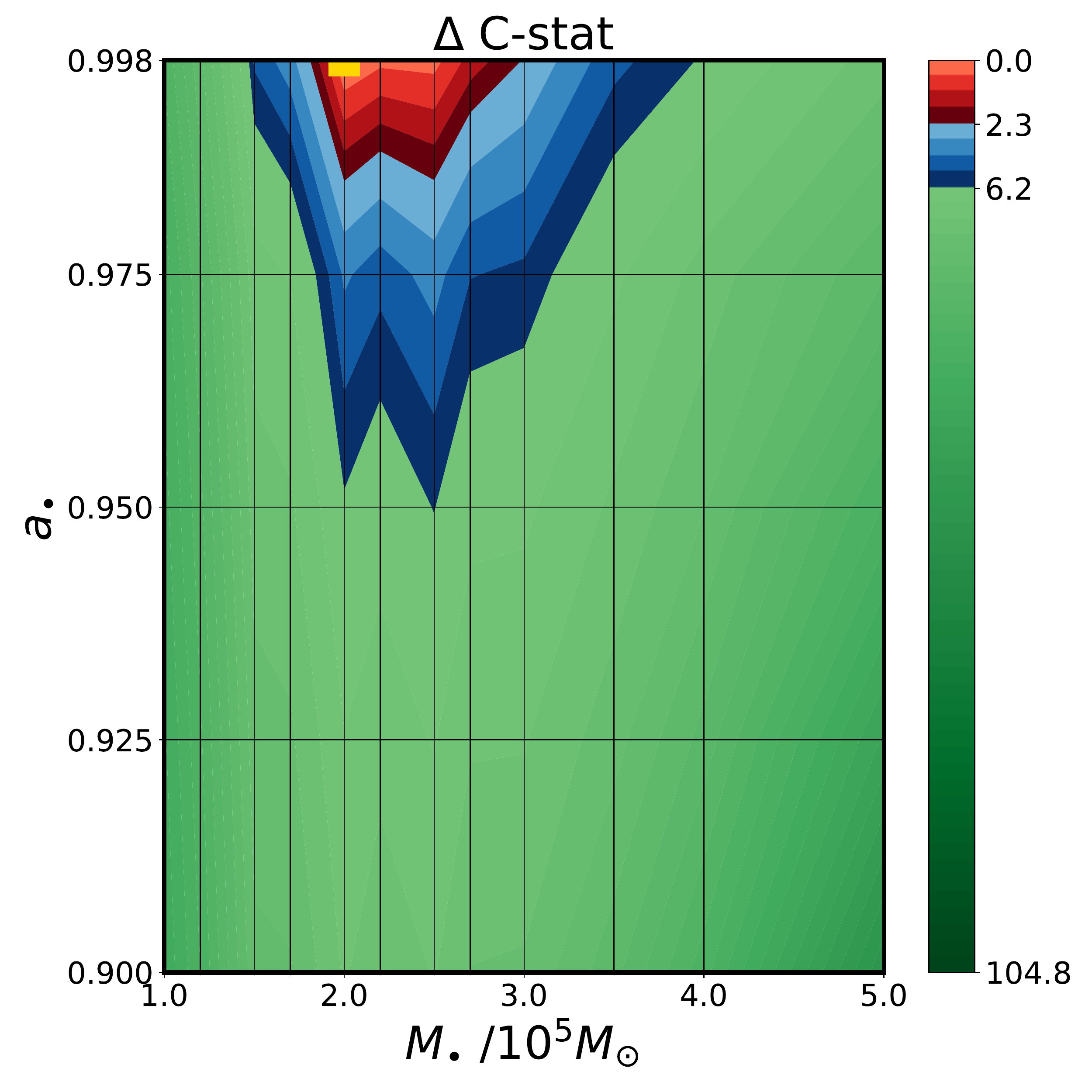}
    \caption{Constraints on $M_{\bullet}$ and $a_{\bullet}$ from the slim disc$+$thermal Comptonisation model-fit to all the observed X-ray spectra. We calculate the $\Delta$C-stat across a model grid in the $\{M_{\bullet}$, $a_{\bullet}\}$ plane (grid points are indicated by vertices of the black lines) and then fill in the colour contours by linear interpolation. The best--fit point with the lowest C-stat is marked by a yellow square. Areas within 1$\sigma$ and 2$\sigma$ are filled by red and blue colours, respectively.  $M_{\bullet}$ and a$_\bullet$ are constrained to be $2.0^{+1.0}_{-0.3}\times10^{5}$~M$_\odot$ and $>0.97$, respectively.}
    \label{fig:contour}
\end{figure}

We find a best fit at $M_{\bullet}=2\times10^5$~M$_\odot$ and $a_{\bullet}=0.998$ for the thermal Comptonisation scenario. We find a total C-stat/d.o.f.~$=2024/2164$. Fig~\ref{fig:contour} shows the $\Delta$C-stat contour in $\{M_{\bullet}$, $a_{\bullet}\}$ space. From the contour, we can constrain the mass of the BH in J150052 to be $2.0^{+1.0}_{-0.3}\times10^{5}$~M$_\odot$; the lower limit to the BH spin is constrained to be $>0.97$ at the 1-$\sigma$ 84.1\% single--sided confidence level. The residuals and the parameter values of the best-fit are summarised in Fig.~\ref{fig:res} and Table~\ref{tb:Cpts}. 

Compared to the slim--disc--alone case when $f_c$ is allowed to have a value larger than 2.4 (section \ref{sc:model}), $\Delta$AIC is 149 for the thermal Comptonisation scenario. In Fig.~\ref{fig:uf_c}, we show the best--fit spectrum of C2--C8 (combined and treated as a single epoch), de--convolved from the detector response curve (i.e., the unfolded spectrum), illustrating the impact of the thermal Comptonisation on the disc continuum. From the residuals of this best--fit (Fig.~\ref{fig:res}) we can see that a slim disc$+$thermal Comptonisation model can describe the observed spectra during all J150052 early epochs well. The source intrinsic absorption $N_{H,i}$ is consistent with being constant in time (from C1 to X6) within 3$\sigma$ errors. The best-fit overall value is $(0.28\pm0.01)\times10^{22}$cm$^{-2}$, in agreement with the previous estimate of $N_{H,i}=(0.26\pm0.06)\times10^{22}$cm$^{-2}$ based on the thin disc assumption \citep[][]{lin2022follow}. Furthermore, compared to the test case when we tie all $\tau$ values, letting $\tau$ free--to--vary between early epochs improves the fit significantly (C-stat/d.o.f.~from 2047/2167 to 2024/2164, $\Delta$AIC$ =17$). While the uncertainties on the best-fit value for $\tau$ is such that no significant trend or changes in its value can be discerned (Table~\ref{tb:Cpts}), the errors can be reduced if we perform the joint--fit with $N_{H,i}$ fixed to the best--fit value of $0.28\times10^{22}$cm$^{-2}$ and $kT_e$ fixed to 2.3~keV. We will discuss the potential change of $\tau$ between epochs as well as the physical origin of the Comptonisation process in the next section.  

Meanwhile, $\dot m$ is estimated to decrease by roughly an order of magnitude over the period the spectra were obtained. In this scenario, the spectral hardening $f_c$ at early epochs is constrained to be $>2.36$, below the upper limit of 2.4. Compared to the slim--disc--only scenario (section \ref{sc:model}), the best-fit $f_c$ obtained at early epochs is more in line with theoretical expectations.

We also investigate if the \{$M_{\bullet}$, $a_{\bullet}$\} constraints are sensitive to the value of $f_c=2.2$ we used for the late-epoch spectral fits.
Freeing the late--time $f_c$ (but fixing it to be constant over the late epochs) results in a best--fit with a slightly higher $M_{\bullet} = 2.7\times10^{5} M_\odot$ (Fig.~\ref{fig:contour-sp}), while the spin value is still consistent with the maximal spin. The late--time $f_c$ is constrained to be $>2.27$ (1$\sigma$ error), at the best--fitted \{$M_{\bullet}$, $a_{\bullet}$\} grid--point. However, by varying the late--time $f_c$, the constrained 1$\sigma$ error range is not changed essentially for either $M_{\bullet}$ or $a_{\bullet}$, and $\Delta$AIC$ = 0$ (C-stat/d.o.f.~$= 2023/2163$). Therefore, we conclude that our choice of $f_c=2.2$ for the late--epoch spectra did not influence the constraints on either $M_{\bullet}$ or $a_{\bullet}$ significantly.

\begin{table*}
\renewcommand{\arraystretch}{1.5}
\centering
\caption[]{Best-fit parameters for the fit function comprised of the slim disc model convolved by a thermal Comptonisation model. The fit function for the source model as used in {\sc XSPEC} is given below, followed by the best-fit $\{M_{\bullet}$, $a_{\bullet}\}$, the constraint on the source inclination, the C-stat/d.o.f.~from the joint-fit to all the spectra, and other epoch-dependent parameter values. Values fixed during the fit are given in between brackets. The accretion rate $\dot m$ is in the unit of the Eddington--limited accretion rate $\dot m_{\rm Edd}=1.37\times10^{21}$~kg~s$^{-1}(M_{\bullet}/10^{6}M_{\odot})$. Note that the actual $\dot M$ in kg~s$^{-1}$ units is not identical to the $\dot m$, and needs to be calculated (see section \ref{sc:model}). The $f_c$ is the spectral hardening parameter. For the thermal Comptonisation model, the Thomson optical depth $\tau$ parameter and the electron temperature $kT_e$ are given, while the covering fraction of \texttt{thcomp} is fixed to unity during the early epoch spectra.} 

\begin{tabular}{ccc|ccccc|c}
\hline
\multicolumn{9}{c}{\texttt{TBabs*zTBabs*thcomp*slimdisc}}\\
\multicolumn{9}{c}{$M_{\bullet}=2\times10^5$~M$_\odot$, $a_{\bullet}=0.998$: $\theta<19^{\circ}$}\\
\multicolumn{9}{c}{C-stat/d.o.f.~$=2024/2164$}\\
&Time since MJD~53574&Epoch&\texttt{zTBabs}&\multicolumn{2}{c}{\texttt{thcomp}}&\multicolumn{2}{c|}{\texttt{slimdisc}}&C-stat/data bins\\
&Days&(in time sequence)&$N_{H,i}$ ($10^{22}cm^{-2}$)&$\tau$&$kT_e$ (keV)&$\dot m$ ($\dot m_{\rm Edd}$)&$f_c$\\
\hline
Early Epoch&1048&C1&$0.25^{+0.05}_{-0.04}$&$4\pm2$&$2.3^{+2.7}_{-0.8}$&$8^{+28}_{-4}$&$>2.36$&180/210\\
&1299&X1&$0.31^{+0.01}_{-0.01}$&$4\pm2$&=C1&$>27$&=C1&339/401\\
&1305&X2&$0.30^{+0.01}_{-0.03}$&$4\pm2$&=C1&$>7.8$&=C1&412/397\\
&2080&C2--C8&$0.23^{+0.02}_{-0.01}$&$5\pm2$&=C1&$1.0\pm0.1$&=C1&386/422\\
\hline
Late Epoch&3502&C9&$0.24^{+0.03}_{-0.03}$&-&-&$1.6^{+0.3}_{-0.2}$&(2.2)&85/66\\
&4381&X3&$0.27^{+0.02}_{-0.02}$&-&-&$1.9^{+0.4}_{-0.3}$&=C9&194/226\\
&4400&X4&$0.24^{+0.03}_{-0.03}$&-&-&$1.5^{+0.5}_{-0.3}$&=C9&80/124\\
&4564&X5&$0.28^{+0.04}_{-0.04}$&-&-&$2.6^{+1.5}_{-0.7}$&=C9&108/118\\
&5326&X6&$0.24^{+0.02}_{-0.02}$&-&-&$1.2^{+0.2}_{-0.1}$&=C9&241/225\\
\hline
\end{tabular}
\label{tb:Cpts}
\end{table*}

\section{Discussion}
\subsection{Implications of IMBH mass and spin}

Fitting the XMM-{\it Newton} and {\it Chandra} X-ray spectra using a slim disc model supplemented by a thermal Comptonisation model to account for the presence of a second, harder spectral component at observing epochs between 2008--2014, we constrain the mass and spin of the BH in J150052 to be $M_{\bullet}=2.0^{+1.0}_{-0.3}\times10^{5}M_\odot$ and $a_{\bullet}>0.97$. The mass uncertainties are given at the 68\% confidence level (1$\sigma$ single parameter), and the limit on the spin is at the 84.1\% confidence level (single--sided 1$\sigma$ single parameter). By modelling all the late--epoch spectra with a physical model for thin disc$+$corona accretion \texttt{optxagnf} (assuming the inclination $\theta=60^{\circ}$; \citealt{done2012intrinsic}), \citet[][]{lin2022follow} estimate the mass of J150052 to be a few~$\times10^{5}$~$M_{\odot}$. Our measurements are consistent with this previous mass estimate, though our best--fit does not include a powerlaw component (as the non--thermal Comptonisation in the corona considered by \texttt{optxagnf}) for the late--epoch spectra. Furthermore, \citet{lin2022follow} find the estimated mass and spin are degenerate and the BH mass would be $\approx7.6\times10^{5}$~$M_{\odot}$ if $a_{\bullet}=0.998$, but we find a smaller BH mass and here we do not see the mass--spin degeneracy (Fig.~\ref{fig:contour}).

To directly compare our results to the estimates from the thin disc$+$corona scenario, we try to jointly fit only the spectra at late epochs with the slim disc model. Through this fit, we also test if our mass and spin constraints are driven mainly by the late--time data. We find that using only the data from late epochs will not change the best--fit \{$M_{\bullet},a_{\bullet}$\} values, and they are insensitive to the inclination given the data quality ($\theta$ is constrained to be $<66^{\circ}$). Meanwhile, the 1~$\sigma$ uncertainty regions for the BH mass and spin increase. We find there is a mass--spin degeneracy similar to that found by \citet{lin2022follow} through their thin disc modelling (a higher mass corresponds to a higher spin; Fig.~\ref{fig:contour-late}). We conclude that, while modelling late--time TDE spectra alone can constrain the mass and the spin well, we need the consistent modelling of both early-- and late--epoch spectra to break the mass--spin degeneracy and minimise uncertainties. This conclusion is in line with the results of \citet[][]{wen2022library}: they find the mass--spin degeneracy of the slim disc solution can be broken when one or more epochs of data obtained when the accretion rate is well above the Eddington--limit are included in the fit.

Despite recovering the mass--spin degeneracy, our joint--fit of only the late--epoch spectra  (Fig.~\ref{fig:contour-late}) still suggests a BH mass lower than $7.6\times10^{5}$~$M_{\odot}$. To check this, we simulate the X-ray spectrum using the best--fit slim disc model from fitting only the late--epoch spectra. For a thin disc\footnote{In practice, an \texttt{optxagnf} model with the corona switched off ($r_{\rm cor}=0$ in \texttt{optxagnf}).} to fit the simulated spectrum, we find either the spin needs to decrease ($\sim0.6$) or the mass needs to increase ($\sim8\times10^{5}$~$M_{\odot}$). This degeneracy appears to be similar to that found by \citet[][]{lin2022follow} when they fit late--epoch spectra using \texttt{optxagnf} that has an additional powerlaw component. Meanwhile, we find the thermal disc emission in \texttt{optxagnf} from \citet[][]{lin2022follow} is similar to our best--fit slim disc model (Fig.~\ref{fig:opt-late}). Thus, we conclude the difference in the mass constraints is mainly due to the degeneracy between the thin-- and the slim-- disc model, while whether the high--energy end of the spectra is modelled by the powerlaw or not has little impact on the disc constraints. As another test, we can fit the data with the slim disc model while forcing the $N_{H,i}=0.26\times10^{22}$cm$^{-2}$, $M_{\bullet}=8\times10^{5}$~$M_{\odot}$, $\theta=60^{\circ}$, and $a_{\bullet}=0.998$ (similar to the best--fit \texttt{optxagnf} disc parameters in \citealt[][]{lin2022follow}). We find the slim disc of such settings fails to fit the spectra (Fig.~\ref{fig:opt-late}), and when adding a powerlaw the best--fit is still significantly worse (C-stat/d.o.f.~$=761/748$) than the one in Fig.~\ref{fig:contour-late}. Our tests show the difference between a thin disc and a slim disc can not be neglect when estimating the BH mass from TDE spectra at Eddington accretion rates.

The slim disc model that we use to constrain the BH mass and spin assumes a relativistic, stationary slim disc accretion disc \citep[][]{abramowicz1988slim,skadowski2009slim,skadowski2011relativistic}. An important assumption in our modelling is that the accretion disc should be aligned with the spinning BH equatorial plane. The angular momentum vector of the orbit of the star that has been disrupted is likely to be inclined with respect to the black hole spin vector, possibly resulting in a tilted accretion disc after the disruption. %The typical time for the stream of stellar debris to circularise and form a (tilted) disc after disruption is $\lesssim10$ days for a BH of 10$^{5}$~$M_{\odot}$ \citep{bonnerot2017long}. This time could be significantly prolonged if nodal precession induced by the rapid BH spin causes the stream of stellar debris to be lifted out of its original orbital plane such that it will avoid self-interaction. However, if that happens the whole TDE electro-magnetic flare is likely to be delayed \citep[e.g.,][]{guillochon2015dark,hayasaki2016circularization}.

%As soon as the stream self-interacts, it will quickly circularize, and a disc will form. Subsequently, the (inner) disc will precess, and, with time, align itself with the equatorial plane of the BH spin. 
The time it takes for a tilted disc to be aligned into the equatorial plane is likely much faster than $10^{2}$~days, %given 
for a $M_{\bullet}\sim10^{5}$~$M_{\odot}$ and a very high spin value
%$a_{\bullet}=0.998$,
due to a combination of the Bardeen-Petterson effect \citep[][]{stone2012observing} and internal torques (which is the dominant mechanism in the case of a fast--spinning BH; \citealt[][]{franchini2016lense}). Because J150052 was first detected in 2006, three years before the observational epoch we labelled as C1, the disc responsible for the X-ray emission is likely to have both circularized \citep[][]{lin2017likely} and aligned itself with the BH equatorial plane by epoch C1. Thus our assumption about a slim disk aligned with the BH spin is 
%likely to be 
%justified.
reasonable.

The only other spin measurement of an IMBH candidate is for the TDE 3XMM~J215022.4-055108 (J2150), where people also find a high spin of $a_{\bullet}\gtrsim0.7$ \citep[e.g.,][]{wen2021mass}. It is not surprising to find IMBH candidates in X-ray selected TDEs. Theory predicts that the rate of TDEs will be dominated by the smallest BH mass range with a high occupation fraction in a dense stellar environment \citep{wang2004revised, stone2016rates} and that, conversely, it will be suppressed for BH masses $M_{\bullet}\gtrsim10^{7.5}$~$M_{\odot}$ \citep[e.g.,][]{kesden2012tidal,stone2016rates}. Because smaller BH masses and higher spin parameters produce brighter soft X-ray emission, (flux-limited) X-ray selected TDE samples will be biased towards this parameter combination \citep{jonker2020implications, mummery2021maximum}. 

The mass and spin measurements of J150052 (and J2150) can shed light on how IMBHs form. Furthermore, an IMBH is likely a key phase in the growth of SMBHs.  
So far, three major classes of theories for the formation of an IMBH have been proposed: runaway collisions of main--sequence stars, which subsequently collapse due to a general relativistic instability (producing IMBH seeds of $M_{\bullet}\sim10^3-10^4$~$M_{\odot}$; e.g., \citealt[][]{zwart2002runaway}); the growth of a seed stellar--mass BH ($M_{\bullet}<10^{2}$~$M_{\odot}$) through the accretion of gas \citep[e.g.,][]{madau2001massive,greif2011simulations}; and the direct collapse of pristine gas clouds in the early Universe \citep[e.g.,][]{loeb1994collapse,bromm2003formation,lodato2006supermassive}. Theoretical arguments suggest that 
only the direct collapse channel is able to produce a BH of $M_{\bullet}\sim10^5$~$M_{\odot}$ at its birth \citep[][]{inayoshi2020assembly}, but the resulting spin is highly uncertain. We note that if the collapsing cloud goes through a supermassive stellar (SMS) phase before collapsing into a BH, a fast--spinning SMS is  found from  simulations. The collapse of this SMS might result in a BH with a high ($a_{\bullet}\sim0.9$) or even extremal ($a_{\bullet}>0.99$) spin \citep[e.g.,][]{reisswig2013formation,inayoshi2014formation}. 

Besides the possibility that J150052 
formed at $\sim10^5$~$M_{\odot}$ with a high spin, the BH could have started at much lower mass.  In that case,
it must have 
%likely 
gained its last $e$-fold in mass through a subsequent accretion episode,
with the final IMBH spin depending heavily on how the accretion took place. If this accretion proceeds through so--called chaotic 
accretion episodes, where the angular momentum vectors of the infalling gas clouds 
are oriented randomly with respect to the BH spin vector, then the IMBH is likely to be spun--down \citep[e.g.,][]{shibata2002collapse,king2008evolution} and not end up with
J150052's fast spin.
%and thus a fast--spinning IMBH as in the case of J150052 is not formed. 
The spin--down effect also applies to the accretion from preceding TDEs of stars on random$/$isotropic orbits \citep{metzger2016wind}. Therefore, our measurements imply that, if J150052 formed at %order--of--magnitude 
a much lower mass, the seed BH grew to its current mass in accretion episodes where the angular momentum vector of the accreted material was aligned with that of the BH spin.%, in such a way that the observed very high spin in J150052 is possible.

\subsection{Origin of the Comptonisation component}
\label{sc:corona}

Our analysis using phenomenological models implies that 
the spectral continuum of J150052 at early epochs can be well approximated by two BB models of different temperatures. Interestingly, several other systems likely to be accreting at a super-Eddington rate have a similar spectral shape (e.g., \citealt{pinto2017ultraluminous,pinto2021xmm} for ULXs; \citealt{kara2018ultrafast} for a TDE). In this paper, we find that the early--epoch spectra of J150052 can not be well-fit by only the slim disc model. Instead, when the slim disc emission is subsequently altered by the effects of inverse-Comptonisation, the spectra at early epochs can be fitted well. 

We compare our spectral fit results with those derived for ULXs to investigate if a similar corona can 
help explain the similarity in the spectral shape and its evolution among super--Eddington accreting sources. To explain the observed spectral shape in ULXs, it is assumed that, in super--Eddington accreting BH-XRB systems, a ``warm'' ($kT_e\sim1$~keV), optically thick ($\tau\gtrsim10$) region of high--energy electrons with a thermal distribution causes inverse Compton scattering of the thermal photons from the disc \citep[e.g.,][]{magdziarz1998spectral,gladstone2009ultraluminous,done2012intrinsic}. This region, i.e., the corona, could be supplied by the disc atmosphere \citep[e.g.,][]{kubota2019modelling}, and it differs from that of a typical, hot ($kT_e\sim100$~keV), optically thin ($\tau<<$1) corona that is usually invoked to be responsible for the power-law continuum in XRBs at low-hard states \citep[e.g.,][]{belloni2010states} and in AGNs. Our results on J150052 for the thermal Comptonisation scenario show that the corona is similar to a ULX ``warm'' corona, and our constraint on the $kT_e$ is consistent with that derived from using the \texttt{nthcomp} model ($kT_e=1.0^{+2.5}_{-0.3}$~keV; \citealt{lin2022follow}). The current uncertainties in the constraints of $\tau$ (Table~\ref{tb:Cpts}) are largely due to the degeneracies between model parameters. When we re--fit all the spectra jointly with $N_{H,i}$ fixed to the best--fit value of $0.28\times10^{22}$cm$^{-2}$ and also $kT_e$ fixed to 2.3~keV, the goodness--of--fit is  C-stat/d.o.f.~$=2047/2174$ while $\Delta$AIC $=-3$ compared to the fit presented in Table~\ref{tb:Cpts}; the $\tau$ is better constrained at each epoch: $4.4\pm0.2$ (C1), $4.6\pm0.2$ (X1), $4.4\pm0.2$ (X2), and $4.9\pm0.1$ (C2--C8). The optical depth increases at C2--C8, which is in agreement with found from the results in \citealt[][]{lin2022follow}.

The current uncertainties in the constraints of $\tau$ (Table 3) are largely due to the degeneracies between model parameters. When we re--fit all the spectra jointly with $N_{H,i}$ fixed to the best--fit value of 0.28E22 /cm$^2$ and also $kT_e$ fixed to 2.3~keV. In this manner, the goodness--of--fit is C-stat/d.o.f.=2047/2174 while $\Delta$AIC = -3 compared to the fit presented in Table 3) while the $\tau$ is better constrained at each epoch: 4.4+/-0.2 (C1), 4.6+/-0.2 (X1), 4.4+/-0.2 (X2), and 4.9+/-0.1 (C2-C8). The optical depth increases at C2-C8, which is in agreement with found from the results in Lin et al. 2022.

Analytical studies show that an optically thick, ``warm'' corona might not emit a fully--thermalised spectrum \citep[e.g.][]{rozanska2015warm}. We can test the impact of the non--thermal Comptonisation effect on our \{$M_{\bullet}$, $a_{\bullet}$\} constraints by replacing the thermal Comptonisation model \texttt{thcomp} with an empirical, very simple, Comptonisation model \texttt{simpl} \citep{steiner2009simple}. The model \texttt{simpl} mimics the up--scattered continuum by a power-law without assuming a specific energy distribution of the electrons. In this manner, we can see how the \{$M_{\bullet}$, $a_{\bullet}$\} constraints 
%are changed 
change if we do not assume a pure, thermal distribution of the coronal electrons. We find that, for this test scenario, the best--fit \{$M_{\bullet}$, $a_{\bullet}$\} values are the same as those from the thermal scenario, with a best--fit C-stat/d.o.f.~$=2020/2164$ ($\Delta$AIC $=4$ compared to the thermal scenario). 
%We find the 
The \{$M_{\bullet}$, $a_{\bullet}$\} constraints are not sensitive to whether the inverse--Comptonisation is done by electrons that have a thermal distribution (Fig.~\ref{fig:contour-simpl}).

In addition to the Comptonisation scenarios, we test whether there could be any disc outflow emission/absorption that can explain the spectral shape observed at early epochs. A disc outflow is seen in simulations of sources at high/super--Eddington accretion rates \citep[e.g.,][]{ohsuga2011global,takeuchi2013clumpy,kitaki2021origins}
%, and this is supported by
and supported by observations \citep[e.g.,][]{middleton2013broad,pinto2016resolved,kara2018ultrafast,pinto2021xmm}. However, there are no high-resolution spectral data of J150052 (e.g., from \xmm~ RGS) with sufficient signal-to-noise ratio to confirm a disc outflow from emission and/or absorption lines.
%the presence of a disc outflow through the presence of emission and/or absorption lines. 
Furthermore, 
using
%we attempted to use 
the atomic library XSTAR in {\sc xspec} \citep[][]{kallman2001photoionization} to model the wind contribution (absorption/emission) to the continuum
%. However, this model 
does not yield a good fit\footnote{The XSTAR model is constructed following \citealt{middleton2013broad}. We assume a clump particle density of $10^{13}$~cm$^{-3}$ and an input ionizing spectrum of a black body with a temperature $kT$ of 0.28 keV and a luminosity of $1\times10^{43}$ erg/s. We then construct an XSTAR grid, stepping between a log($\xi$) of 3 and 5 in 10 linear steps and a column density of $1\times10^{20}$ and $1\times10^{23}$ in 20 logarithmic steps.}. 

%We note, however, that 
It is still possible that an outflow is present in J150052 and that it might contribute to the aforementioned inverse--Comptonisation process. In this case, 
%the absence of the detection of 
not detecting a direct outflow signature could either be due to the lack of high-resolution X-ray
spectra 
%spectroscopic observations 
or 
%because we are looking at 
to observing the disc at a low inclination angle ($\theta\lesssim30^{\circ}$). The low inclination angle could imply that not much of the wind outflow is 
%present 
along the line--of--sight \citep[][]{pinto2017ultraluminous,dai2018unified}. Alternatively, the outflow velocity could be high, which, %especially 
when combined with a low equivalent width of the line features, would lead to broadened lines that are difficult to detect.
If such a wind were present,
we can roughly estimate the associated mass loss.
Based on the simulations of \citet{takeuchi2013clumpy}, we expect wind clump sizes of $\sim10$~$R_g$. Using the $\tau$ and $kT_e$ constrained by our thermal Comptonisation models, we estimate a yearly mass loss of $\sim2\times10^{-3}$~$M_{\odot}$,
%in a wind were it present,
assuming that the wind velocity equals the escape velocity at 10~$R_g$. 

\subsection{Transition from super-- to sub--Eddington accretion}

\begin{figure}
    \centering
    \includegraphics[width=\linewidth]{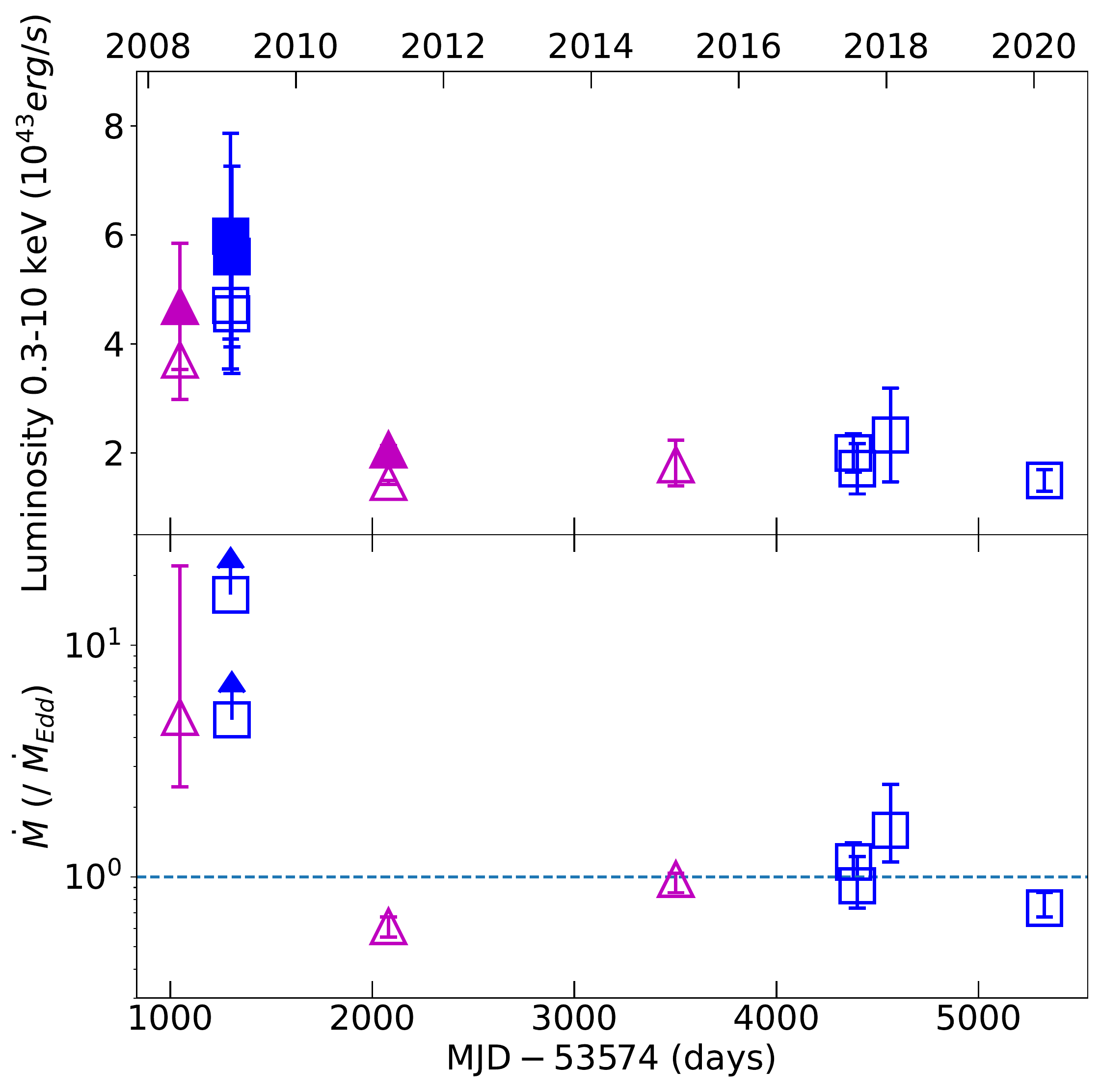}
    \caption{\textit{Upper panel}: Lightcurve of J150052 starting from 2008, constructed from the joint--fits of the slim disc$+$thermal Comptonisation scenario. The format follows Fig.~\ref{fig:lcbb}. The open symbols show the slim disc luminosity, and the filled symbols at early epochs show the total (slim disc$+$Comptonisation) luminosity. The lightcurve is corrected for both the effects of intrinsic and Galactic absorption.
    %Due to computational intensities, we give 
    We show the statistical uncertainty on 
    %the luminosity 
    each luminosity at the 1$\sigma$ (68\%) confidence level. These
    uncertainties are calculated by varying the normalisation of the source fit function and are likely to underestimate those determined when 
    %. The uncertainty determined in this way typically under--estimates the uncertainty in the luminosity calculated when 
    all parameters are allowed to vary. However, the latter is computationally unfeasible in this case. \textit{Bottom panels}: Accretion rate $\dot M$ derived from the slim disc$+$thermal Comptonisation model, in units of the Eddington accretion rate $\dot M_{\rm Edd}=16L_{\rm Edd}/c^2$ (for consistency with \citealt[][]{skadowski2011relativistic,abramowicz2013foundations}). Note that due to this different definition of the Eddington accretion rate from the $\dot m_{\rm Edd}$ in the slim disc %\texttt{slimfit}
    model, the actual $\dot M$ in $\dot M_{\rm Edd}$ units is not identical to the $\dot m$ parameter (in $\dot m_{\rm Edd}$ units) in the model, and needs to be calculated (see section \ref{sc:model}). We use the upper--case $\dot M$ and lower--case $\dot m$ to distinguish them. The Eddington rate $\dot M={\dot M}_{\rm Edd}$ is shown by a dashed horizontal line.}
    %The dashed black curve shows a $\dot M\propto~t^{-\alpha}$ curve fitted to the data, with the best-fit $\alpha=0.5\pm0.1$.}
    \label{fig:mdot}
\end{figure}

Fig.~\ref{fig:mdot} shows the change of the source luminosity and the slim disc accretion rate $\dot M$ over time based on our best-fit slim disc+thermal Comptonisation model, assuming a distance of $\sim$690 Mpc \citep[][]{lin2017likely}. Note that the actual $\dot M$ in kg~s$^{-1}$ units is not identical to the $\dot m$ parameter (in $\dot m_{\rm Edd}$ units) in the model, and needs to be calculated ($\dot M=\dot m\cdot\dot m_{\rm Edd}$; see Section \ref{sc:model}). We use the upper--case $\dot M$ to distinguish from the $\dot m$. %We find that 

During the long-term decay of J150052, $\dot M$ decreases from $\sim10$~$\dot M_{\rm Edd}$ to $\lesssim1$~$\dot M_{\rm Edd}$ (here $\dot M_{\rm Edd}=16L_{\rm Edd}/c^2$ for consistency with \citealt{skadowski2011relativistic,abramowicz2013foundations}). The decrease in $\dot M$ might lead to a decreased amount of the inverse--Comptonisation if fewer high-energy electrons are generated at lower mass accretion rates (e.g., if the wind becomes less powerful as the mass accretion rate decreases below the Eddington limit). Therefore, it is possible that, as $\dot M$ decreases, the source spectrum can be well described by the slim disc model without the thermal Comptonisation component.
%We can infer the total mass $M_{\rm acc}$ accreted by the BH from C1 to X6 by integrating $\dot M$ over time. We fit a $\dot M\propto~t^{-\alpha}$ function to approximate the long--term lightcurve (Fig.~\ref{fig:mdot}), obtaining the best--fit $\alpha=0.5\pm0.1$. We find that the BH has accreted $\sim0.12$~$M_{\odot}$ from C1 to X6, 
%%and this is 
%%in agreement with the scenario of 
%consistent with an IMBH disrupting a low--mass main--sequence star as proposed by \citet{lin2022follow} based on lightcurve modelling.
The previous studies by \citet{lin2017likely,lin2022follow} already showed that the spectra at late epochs are different from the spectra preceding C9, and they proposed that the spectral evolution is caused by the transition from a super--Eddington to a sub--Eddington accretion rate. Our results are in agreement with this suggestion.

Spectral analyses of other TDEs indicate that TDEs might undergo a spectral transition during the decay \citep[e.g.,][]{bade1996detection,komossa2004huge,wevers2019black,jonker2020implications,wevers2021rapid}. Several TDEs %are shown 
appear to transition from a soft, disc--dominated state to a hard, non--thermal state, where the X-ray spectrum is characterised by a power-law. Interestingly, in J150052, we find evidence for a spectral state transition analogous to a transition from the ultra--luminous state of BH--XRBs \citep[e.g.,][]{gladstone2009ultraluminous,motta2012discovery,sutton2013ultraluminous} to the soft state. If the mass accretion rate in J150052 continues to decrease, we predict that the spectrum will exhibit a Comptonised component again when the BH transitions from the soft to the hard state, in analogy to the soft--to--hard state transition in BH--XRBs.

%\textbf{We notice that it seems the source accretion rate $\dot M$ is not monotonically decreasing around Epoch C2--C8. This might be the case when the corona is turning off and the assumption of a steady corona in our model breaks down. Indeed if we un--tie the electron temperature between C2--C8 and other early epochs, the electron temperature at C2--C8 will increase significantly ($kT_e\sim100$~keV, $\tau<1$; see Table \ref{tb:test}). In that case the $\dot M$ will be $\sim1.5$~$\dot M_{\rm Edd}$, more in line with the picture of a gradually--decaying accretion. However, $\Delta$AIC $=4$ for one more free parameter, and thus we do not treat it as an improvement to the joint--fit.}

\subsection{Constraining ultralight boson masses} 

Following the procedure described in \citet{wen2021mass}, we derive constraints on the mass of hypothetical ultralight bosons based on our spin and mass measurements of J150052. As mentioned in \S \ref{sec:intro}, a BH will only spin down efficiently when bosons of mass $m$ exist such that $\frac{GM_{\bullet}m}{c\hbar}\sim1$ (here $\hbar$ is the reduced Planck constant; \citealt[][]{bardeen1972rotating,press1972floating}). For $\frac{GM_{\bullet}m}{c\hbar}$ values away from 1, the timescale $\tau_{\rm I}$ for growing the instability increases exponentially. We compute $\tau_{\rm I}$ as a function of $m$ to investigate the ultralight boson constraints that can be imposed from J150052. 

\begin{figure}
    \centering
    %\subfloat{\includegraphics[width=.9\linewidth]{figures/bosons3.pdf}}\hfill
    \subfloat{\includegraphics[width=.9\linewidth]{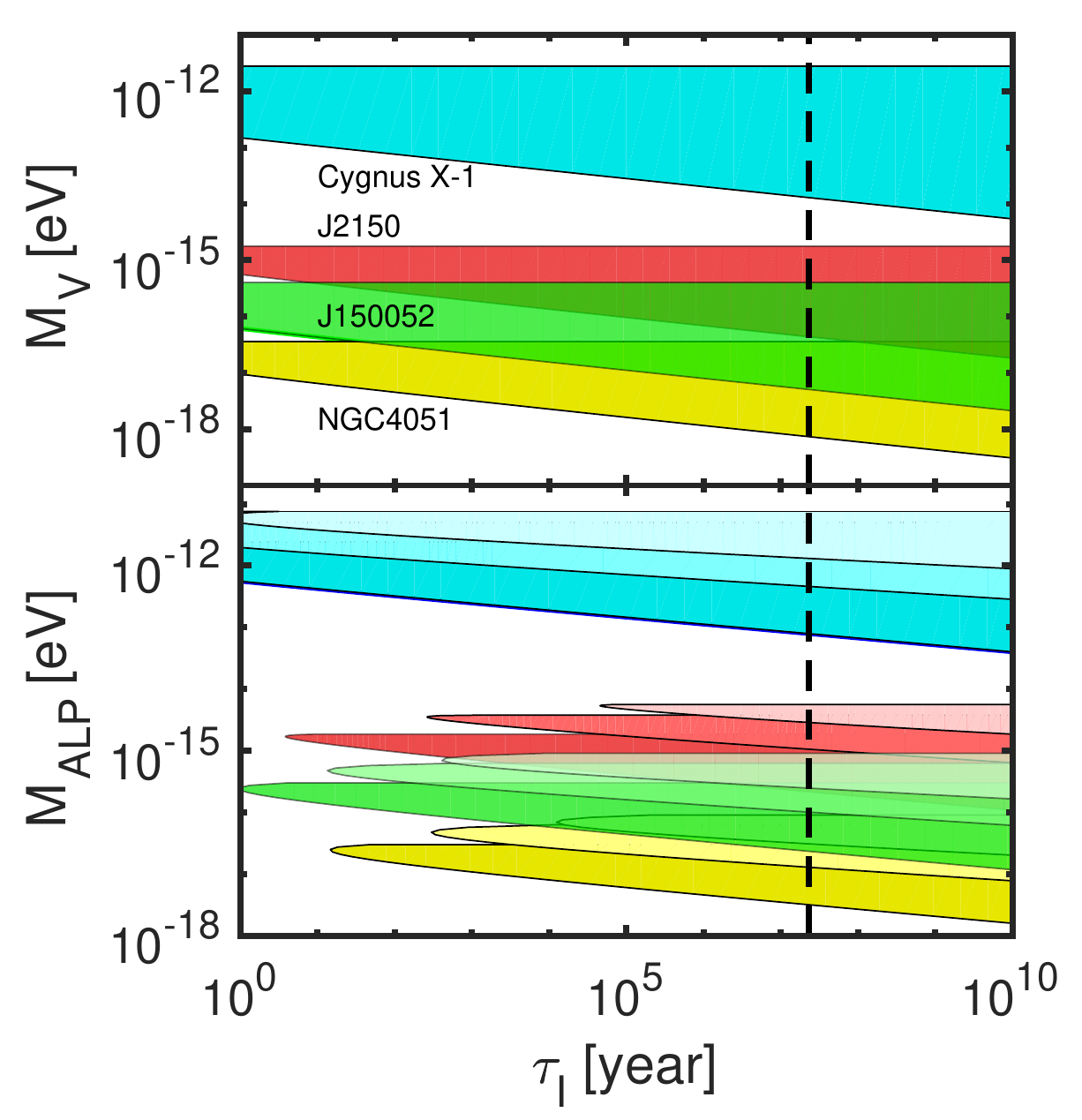}}\hfill
    %\subfloat{\includegraphics[width=.9\linewidth]{figures/bosons3.pdf}}\hfill
    \caption{Exclusion ranges for the masses of hypothetical ultralight bosons, including both Proca vector bosons (mass $M_{\rm V}$; top panel) and scalar axion--like particles (mass $M_{\rm ALP}$; bottom panel), as a function of the BH
    instability timescale $\tau_I$. The format follows Fig.~5 in \citet{wen2021mass}, who derive the exclusion regions from the BH mass and spin measurements of 3XMM J215022.4--055108 (J2150; red contours). We add the exclusion regions derived from our J150052 measurements (green). The contours denote the excluded masses for a given \{$M_{\bullet}$, $a_{\bullet}$\} pair and %instability time 
    $\tau_I$ \citep[][]{wen2021mass}. The cyan and yellow contours are the constraints 
    from the stellar BH system Cygnus~X-1 and the supermassive BH system NGC~4051, respectively. The dark, light, and lighter contours in the lower panel denote the cases of low-order instability modes (mode--number $m=$1, 2 and 3, respectively). The dashed vertical line indicates the timescale
    ($\sim 3\times 10^7$~yr) for a BH with $a_{\bullet}=$ 0 to undergo an Eddington--limited $e$--fold in mass growth, which is the shortest timescale over which significant astrophysical spin--up will occur under the Eddington--limit (the Salpeter timescale; \citealt[][]{salpeter1964accretion}). If the instability timescale is shorter than this spin--up timescale, the BH will never reach a high spin. Thus,
    unless super--Eddington accretion is dominating the mass growth,
    our discovery of J150052's rapid spin rules out the short instability timescales to the left side of the vertical line and allows the wider range of excluded particle masses to the right.
    %, unless super--Eddington accretion is dominating the mass growth.
    }
    \label{fig:axion}
\end{figure}

Fig.~\ref{fig:axion} shows the excluded particle masses (green coloured regions) derived from our mass and spin measurements of J150052. The black hole instability time on the x--axis is calculated using eqs.~$2.13$ and $2.18$ of \citealt{cardoso2018constraining}, taking our best-fit values for $M_{\bullet}$ and $a_{\bullet}$. We also 
show the constraints derived from other accreting BHs with spin measurements: Cygnus~X-1 ($M_{\bullet}=21.2$~$M_{\odot}$, $a_{\bullet}=0.998$; \citealt{miller2021cygnus}, \citealt{zhao2021re}), NGC~4051 ($M_{\bullet}=1.91\times10^6$~$M_{\odot}$, $a_{\bullet}=0.99$; \citealt{denney2009revised}, \citealt{patrick2012suzaku}), and 3XMM J215022.4-055108 (J2150; $M_\bullet=1.75^{+0.45}_{-0.05}\times10^{4}$~$M_{\odot}$ and $a_{\bullet}=0.80^{+0.12}_{-0.02}$; \citealt{wen2021mass}).

The stellar--mass BH Cygnus~X-1 (cyan regions) excludes massive vector (e.g., dark photons) and scalar fields (e.g., axion--like particles) with $m\sim10^{-12}$~eV, whereas the supermassive BH NGC 4051 (yellow) excludes those particles at $m\sim10^{-17}$~eV. Compared to J2150 (red), which excludes the $m\sim10^{-15}$~eV region, J150052 excludes a new region, $m\sim10^{-16}$~eV, for both kinds of particles. For each \{$M_{\bullet}$, $a_{\bullet}$\} pair measurement, we are able to exclude roughly one to two orders of magnitude in ultralight boson mass across large $\tau_{\rm I}$ range.

As seen in Fig~\ref{fig:axion}, the range of excluded ultralight boson masses increases with $\tau_{\rm I}$. If we further assume Eddington--limited accretion for BHs, it is possible to use timescale arguments to
restrict $\tau_{\rm I}$ to $\gtrsim10^7$~yr, given the high spins of these BHs, and thereby focus on the wider particle mass exclusion ranges 
on the right in Fig.~\ref{fig:axion}. The dashed vertical line in Fig. \ref{fig:axion} shows the Salpeter timescale of $\sim 3\times 10^7$~yr required to spin up a BH from
$a_\bullet=0$ to 1 \citep[][]{salpeter1964accretion}. As this is the timescale for a BH to gain an $e$--fold in mass growth when accreting at the Eddington--limit, it is the shortest timescale over which significant astrophysical spin--up will occur. If the instability timescale
is shorter than this spin--up timescale, the BH will never reach a high spin. Therefore, J150052's high spin and those of the other plotted BHs rule out the short instability timescales to the left of the vertical line, unless super--Eddington accretion is dominating the mass growth.

\section{Conclusions}

In this paper, we present and analyse all the publicly available \xmm{} and \chan{} data of the tidal disruption event J150052 since 2008, obtained during its decade--long decay. We fit the X-ray spectra with the slim disc model \citep[][]{wen2020continuum,wen2021mass}, thereby constraining the black hole mass and spin to a higher precision than previously possible. We have examined the implications of these measurements for the existence of IMBHs and their growth, for the masses of
%the possible existence of
hypothetical ultralight bosons, % and scalar axion particles, as well as 
and for the %possible 
origin
of the observed coronal emission and its evolution.
This analysis is only the second of its kind for an IMBH TDE candidate.
Our conclusions are:

\begin{itemize}

\item The BH mass is $\approx 2\times 10^5$~M$_\odot$. More precisely,
if the coronal emission at early epochs arises from thermal Comptonisation, we obtain $M_{\bullet}=2.0^{+1.0}_{-0.3}\times10^{5}M_\odot$. Here the errors are at the 68\% confidence level. The strong mass constraint demonstrates the potential of using X-ray TDEs to search for IMBHs and is consistent with the previous estimate of a few~$\times10^{5}$~$M_{\odot}$ based on the \texttt{optxagnf} model \citep[][]{lin2022follow}.

\item The lower limit on the BH spin is $>0.97$ at the 1$\sigma$ 84.1\% single--sided confidence level for the slim disc$+$thermal Comptonisation models.
Thus, J150052 is a fast spinning, perhaps near--extremal, IMBH. We discuss different IMBH formation channels; our mass and spin measurements imply that, if J150052 did not form near its current mass ($\sim10^5$~M$_\odot$), then it must have accreted up to its current mass in episodes where the angular momentum vectors of the spin and accreted material were aligned. 

\item Our mass and spin measurements of J150052 rule out both vector bosons and axions of masses $\sim10^{-16}$~eV. Vector bosons and axion--like particles are of astrophysical interest as particle dark matter candidates. We show here that, for the mass and spin pair measurement of J150052, the masses of such ultralight bosons can be significantly constrained.

\item Our spectral analyses suggest that J150052 undergoes a transition during its decay, quenching the corona while the mass accretion rate decreases from super--Eddington to $\approx$Eddington levels. The spectral changes are reminiscent of the state transitions in Galactic ultra--luminous X-ray sources. We discuss the origin of the corona. From the spectral constraints of the Compton component, we infer the corona of J150052 to be optically thick and warm ($kT_e=2.3^{+2.7}_{-0.8}$~keV).
%could be dominated by electron--positron pairs, like the coronae of many AGN.

\end{itemize}

%In this paper, 
By constraining the mass and the spin of J150052, we have demonstrated the potential of using %spectral analyses
the X--ray spectra of TDEs to find IMBHs and measure their masses and spins. Similar analyses of large samples of TDEs with suitable early and/or multi--epoch X--ray observations will ultimately constrain the distributions of BH masses and spins, leading to a better understanding of the formation and evolution of both IMBHs and SMBHs. While such analyses are beyond the scope of this paper, we plan to carry them out in future work.

\section*{Acknowledgements}

We thank the anonymous referee for insightful comments. AIZ and SW thank the UA Department of Astronomy and Steward Observatory for support.
AIZ acknowledges additional funding from NASA ADAP grant \#80NSSC21K0988.
This work used the Dutch national e-infrastructure with the support of the SURF Cooperative using grant no. EINF-1077.

%%%%%%%%%%%%%%%%%%%%%%%%%%%%%%%%%%%%%%%%%%%%%%%%%%
\section*{Data Availability}

All the X--ray data in this paper are publicly available from the data archive of HEASARC (https://heasarc.gsfc.nasa.gov/). A reproduction package is available at DOI: 10.5281/zenodo.6621932.

%%%%%%%%%%%%%%%%%%%% REFERENCES %%%%%%%%%%%%%%%%%%

% The best way to enter references is to use BibTeX:

\bibliographystyle{mnras}
\bibliography{references} % if your bibtex file is called example.bib

% Alternatively you could enter them by hand, like this:
% This method is tedious and prone to error if you have lots of references
%\begin{thebibliography}{99}
%\bibitem[\protect\citeauthoryear{Author}{2012}]{Author2012}
%Author A.~N., 2013, Journal of Improbable Astronomy, 1, 1
%\bibitem[\protect\citeauthoryear{Others}{2013}]{Others2013}
%Others S., 2012, Journal of Interesting Stuff, 17, 198
%\end{thebibliography}

%%%%%%%%%%%%%%%%%%%%%%%%%%%%%%%%%%%%%%%%%%%%%%%%%%

%%%%%%%%%%%%%%%%% APPENDICES %%%%%%%%%%%%%%%%%%%%%
\newpage
\appendix
\section{Supplementary figures}
\begin{figure*}
    \centering
    \subfloat[\label{fig:1bbr}]{\includegraphics[width=0.5\linewidth]{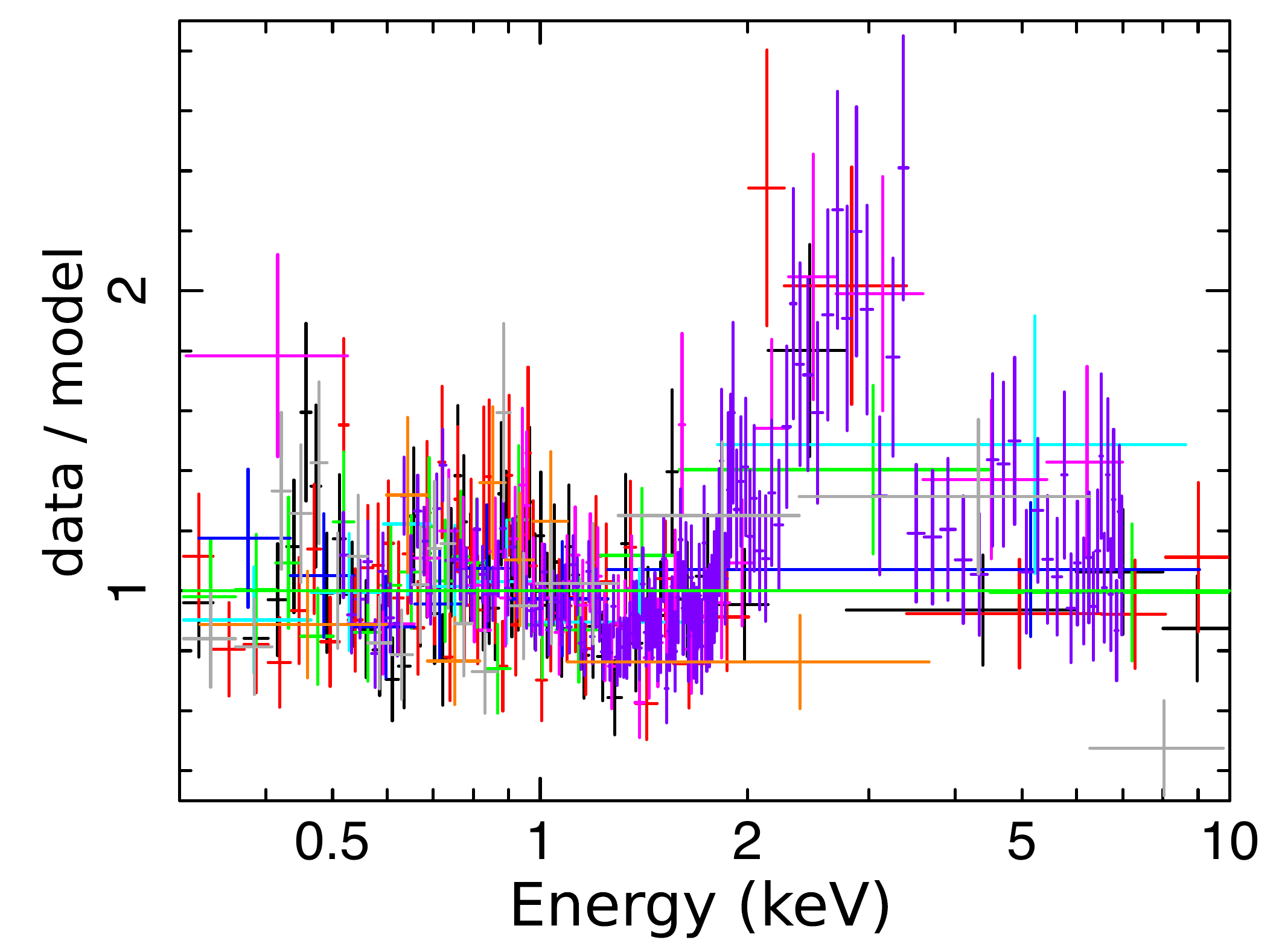}}\hfill
    \subfloat[\label{fig:2bbr}]{\includegraphics[width=0.5\linewidth]{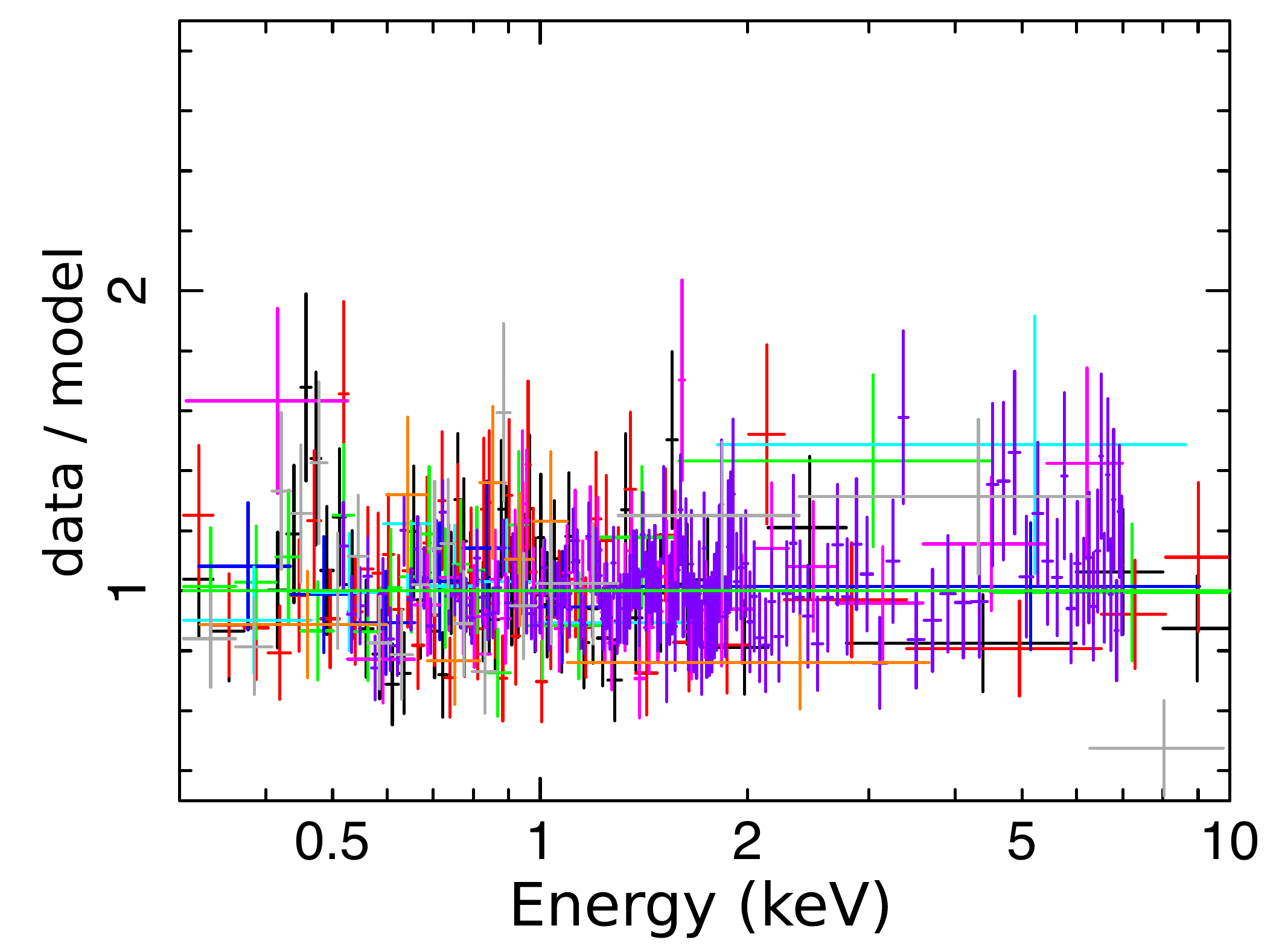}}\hfill
    \caption{Data$/$model ratio for the spectral fits of each epochs. The source fit function is comprised of \textbf{a}) a single BB, or \textbf{b}) two BBs. The source$+$background data at C1, X1, X2, C2--C8, C9, X3, X4, X5, and X6 (in time sequence), are shown in magenta, black, red, purple, orange, green, blue, cyan, and grey, respectively. The background dominates above 3~keV in all epochs. We can see that a single--BB model does not describe the spectra well around 2~keV in several early epochs.}
    \label{fig:bbr}
\end{figure*}

\begin{figure*}
    \centering
    \subfloat[\label{fig:fcall_e}]{\includegraphics[width=0.5\textwidth]{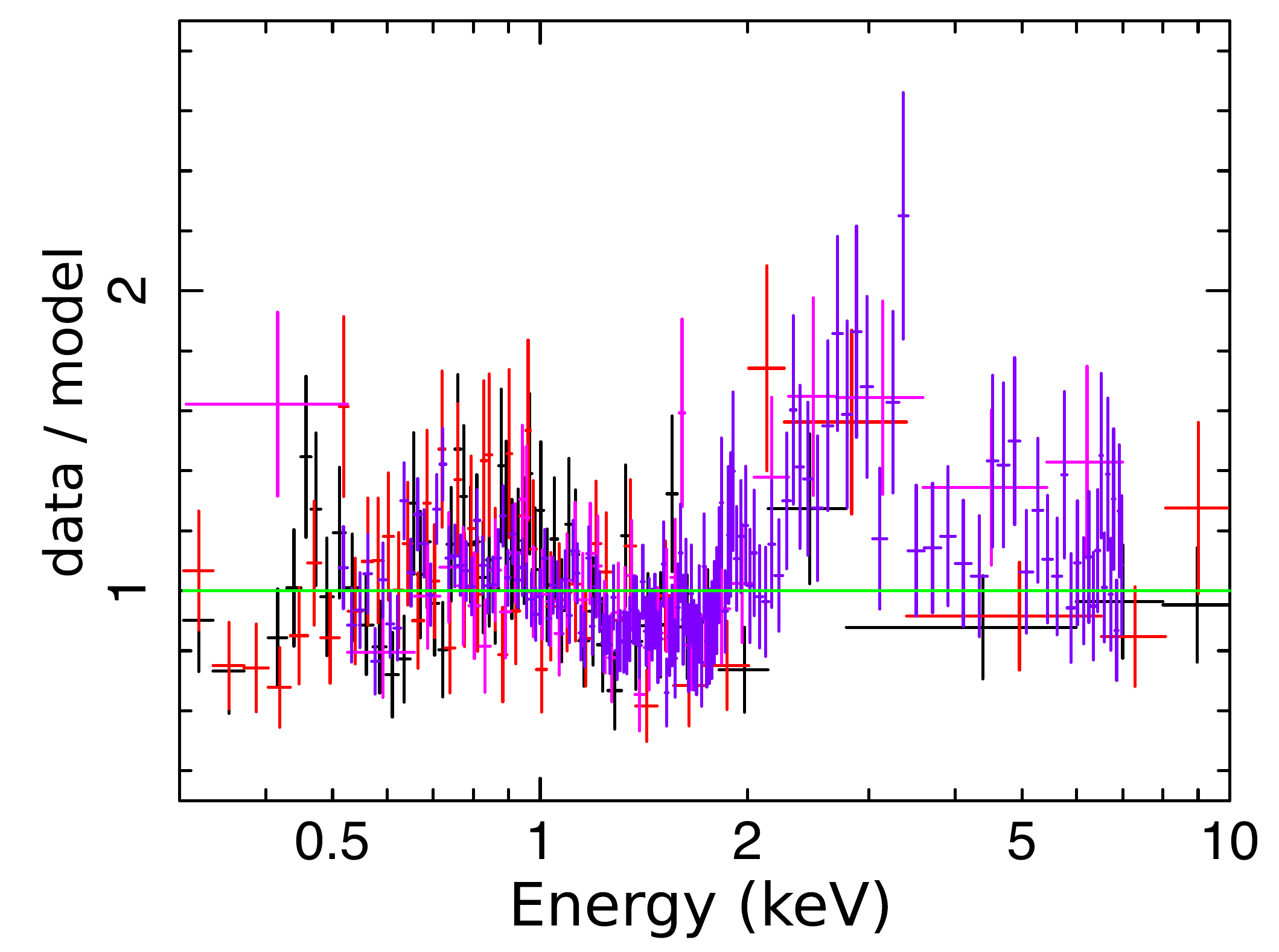}}\hfill
    \subfloat[\label{fig:fcall_l}]{\includegraphics[width=0.5\textwidth]{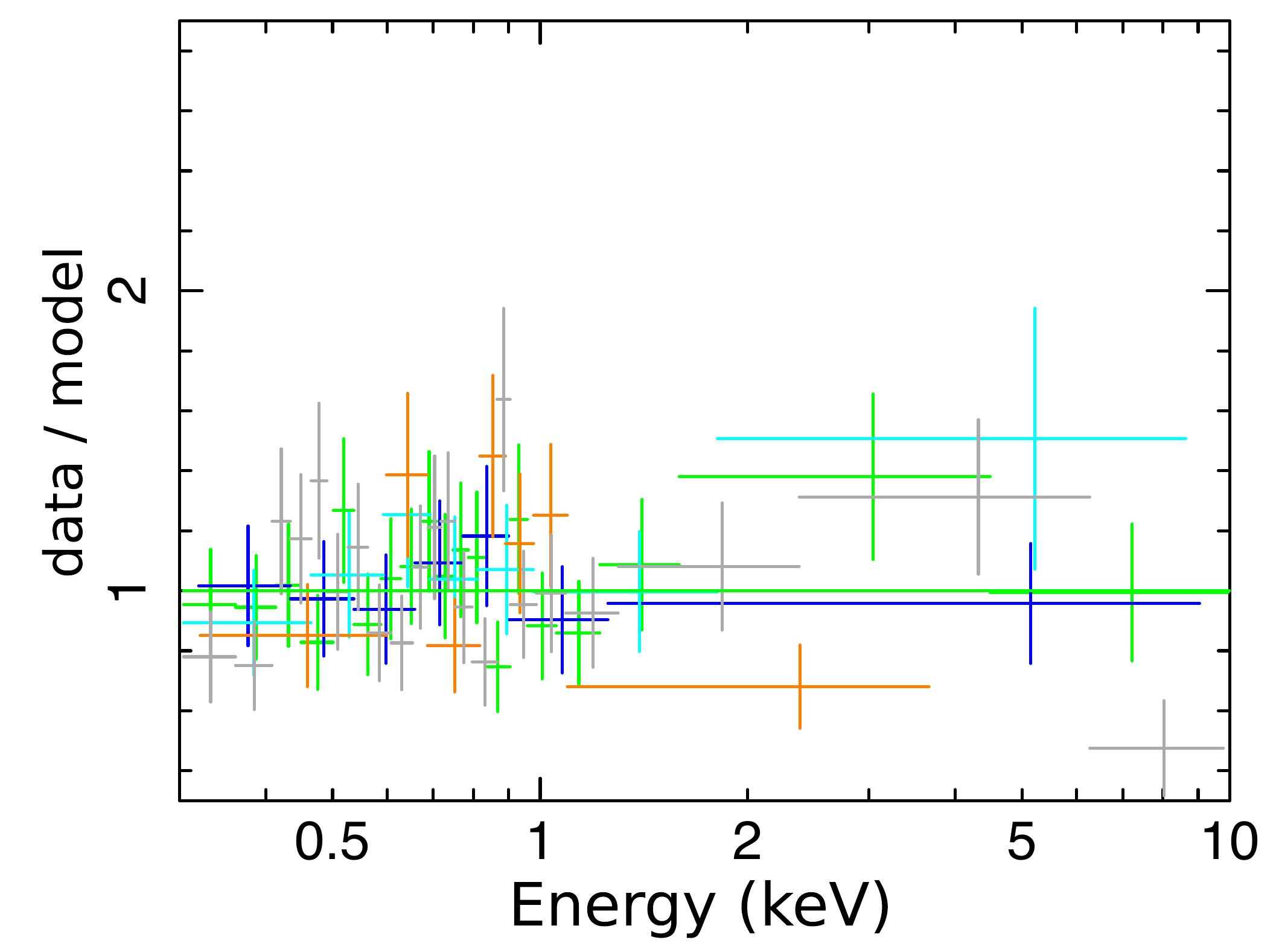}}
    \caption{Data$/$model ratio of the joint-fit of J150052 spectra at \textbf{a)} early epochs and \textbf{b)} late epochs, using a fit function \texttt{TBabs*zTBabs*slimdisc} to describe the source spectra. The colours represent the same as those in Fig.~\ref{fig:bbr}. During the fit, we let the spectral hardening factor $f_c$ free to vary during the early epochs while we fix it to 2.2 during the late epochs. The best-fit $f_c$ during the early epochs is found to be larger than 4.0, whereas in theory $f_c$ should have a maximum value of around 2.4 in the super-Eddington regime \citep[][]{davis2006testing}. Still, strong residuals between 2--3 keV are present.}
    \label{fig:fcall}
\end{figure*}

\begin{figure*}
    \centering
    \subfloat[\label{fig:fc24_e}]{\includegraphics[width=0.5\textwidth]{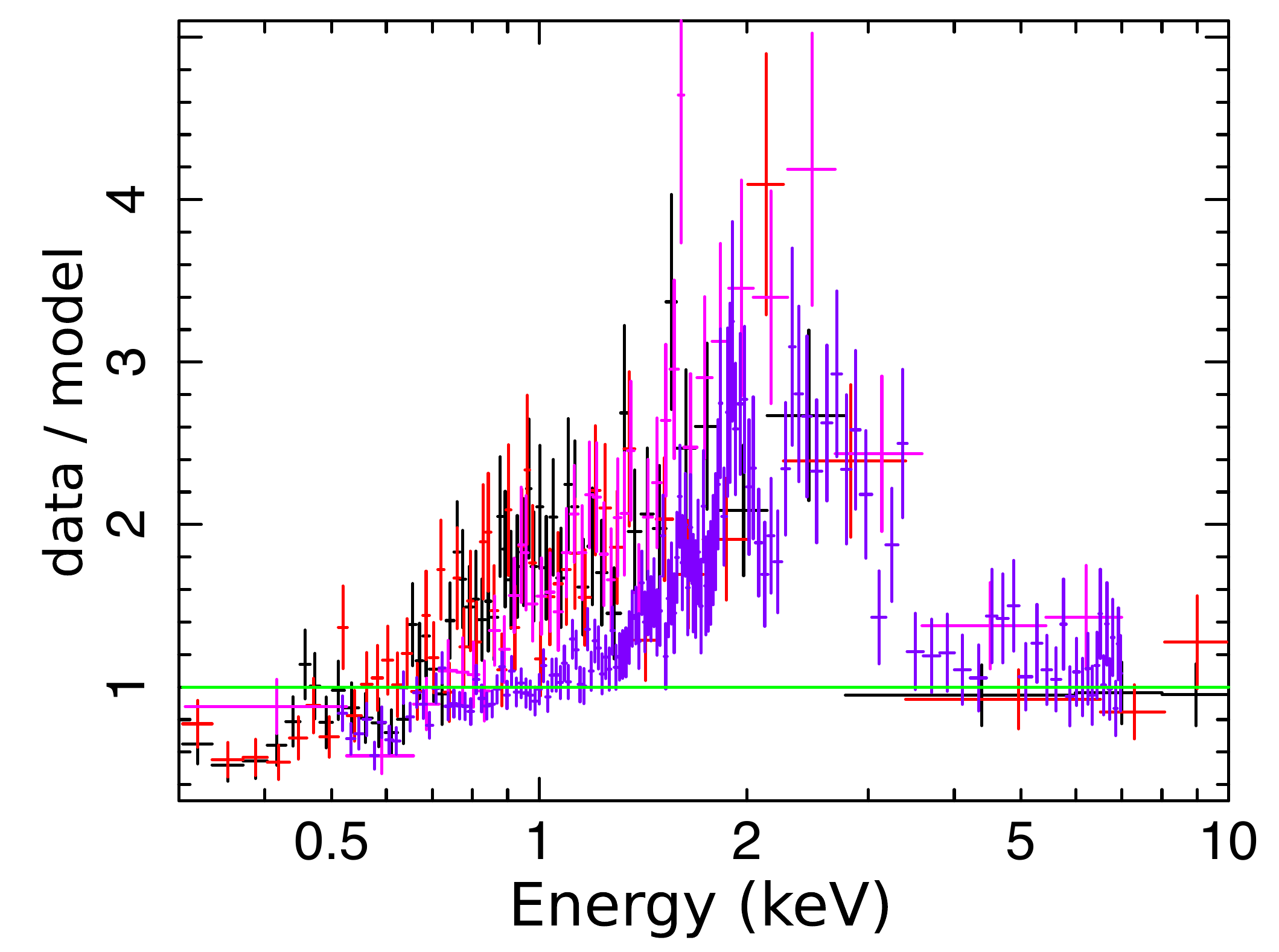}}\hfill
    \subfloat[\label{fig:fc24_l}]{\includegraphics[width=0.5\textwidth]{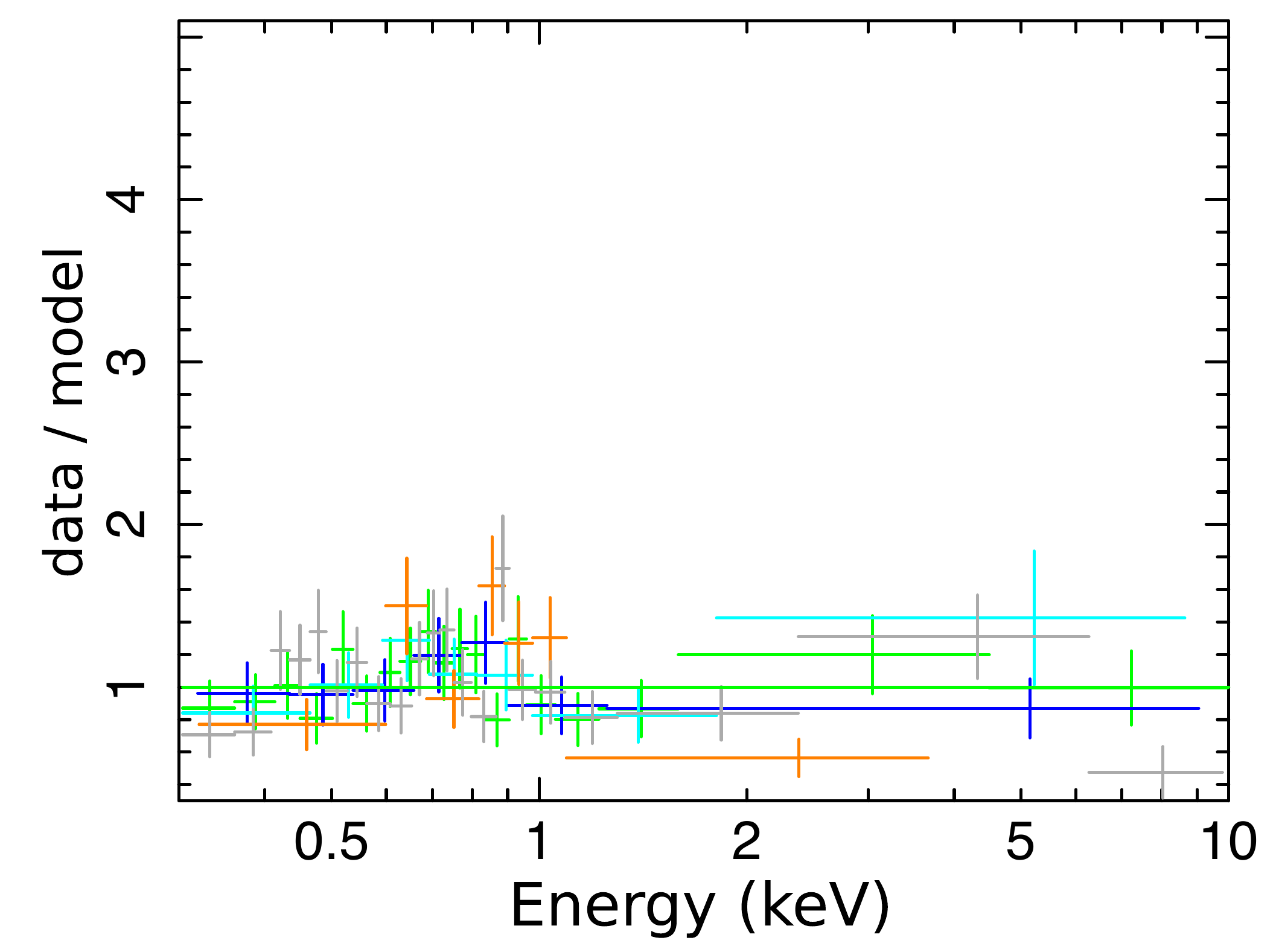}}
    \caption{The same as Fig.~\ref{fig:fcall} but with the $f_c$ parameter fixed to 2.4 for the early epochs and to 2.2 for the late epochs. The excess around 2~keV at early epochs is more prominent than the fit where $f_c$ is left to float freely, showing that the slim disc model can not account for all the hard photons detected at early epochs.}
    \label{fig:fc24}
\end{figure*}

\begin{figure*}
    \centering
    \subfloat[\label{fig:res_s_e}]{\includegraphics[width=0.5\textwidth]{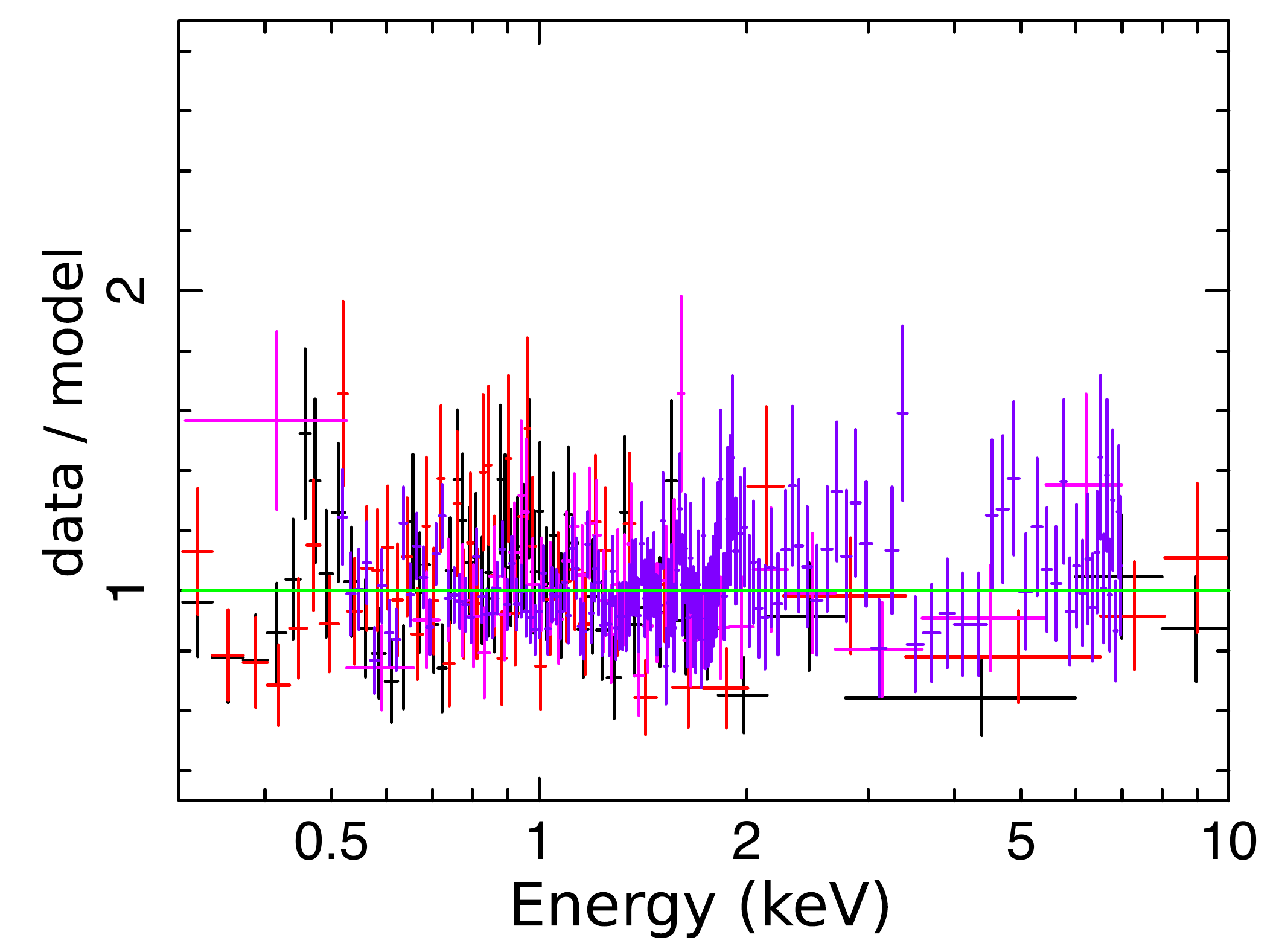}}\hfill
    \subfloat[\label{fig:res_s_l}]{\includegraphics[width=0.5\textwidth]{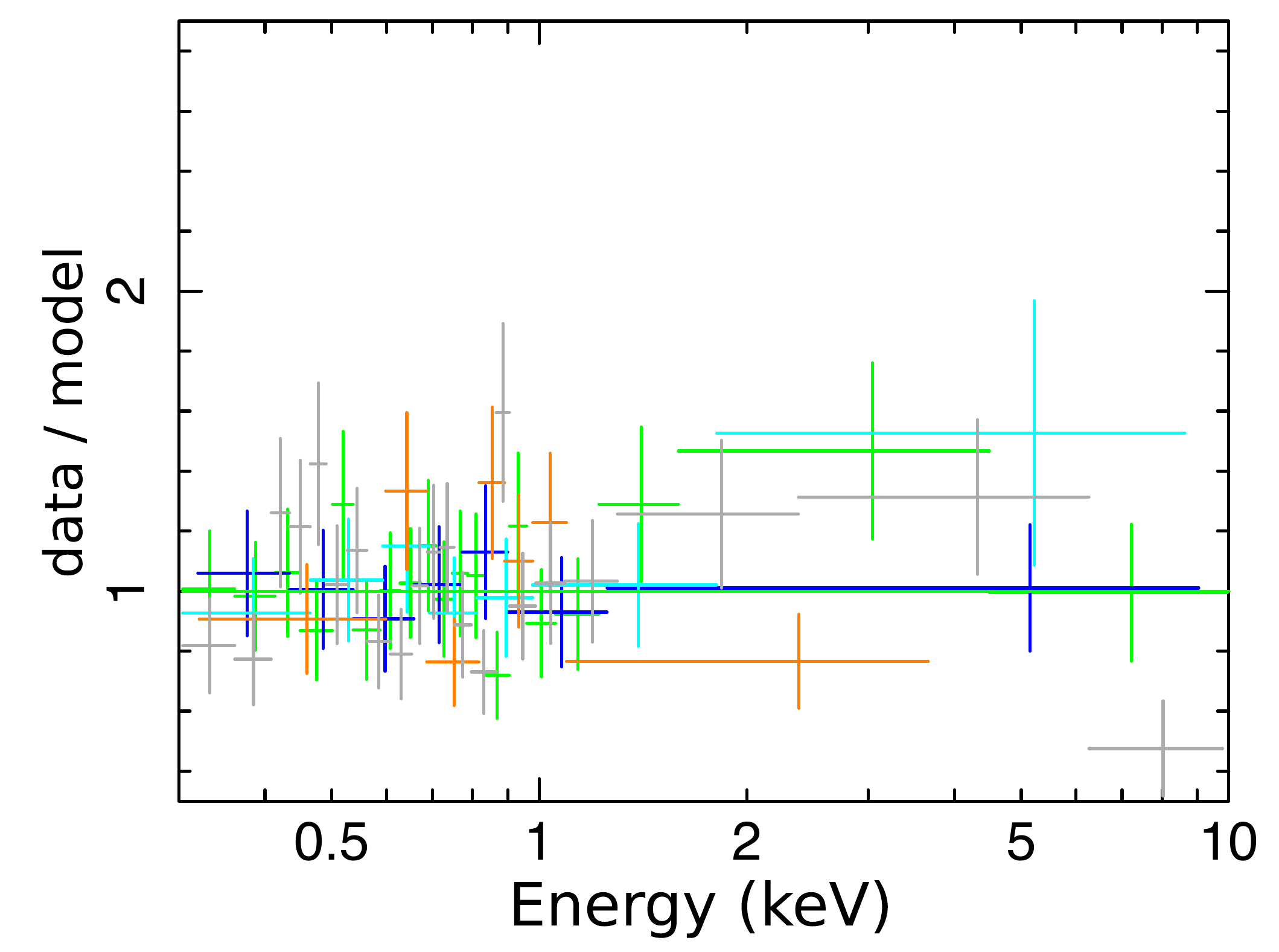}}\hfill
    %\subfloat[\label{fig:res_h_e}]{\includegraphics[width=0.5\textwidth]{figures/res_h_e.pdf}}\hfill
    %\subfloat[\label{fig:res_h_l}]{\includegraphics[width=0.5\textwidth]{figures/res_h_l.pdf}}\hfill
    \caption{Data/model ratio for the joint-fit of the J150052 spectra for the thermal Comptonisation scenario, at \textbf{a)} early epochs and \textbf{b)} late epochs. The colour scheme follows that of Fig.~\ref{fig:fcall}. Compared to the slim--disc--alone scenario (Fig.~\ref{fig:fc24}), adding a Comptonisation component explains the excess flux present in the spectra, particularly around 2~keV at early epochs. During the joint fits we switch off the Comptonisation in the spectra at the late epochs, as those spectra are consistent with the slim disc spectra.}
    \label{fig:res}
\end{figure*}

\begin{figure*}
    \centering
    \includegraphics[width=.5\linewidth]{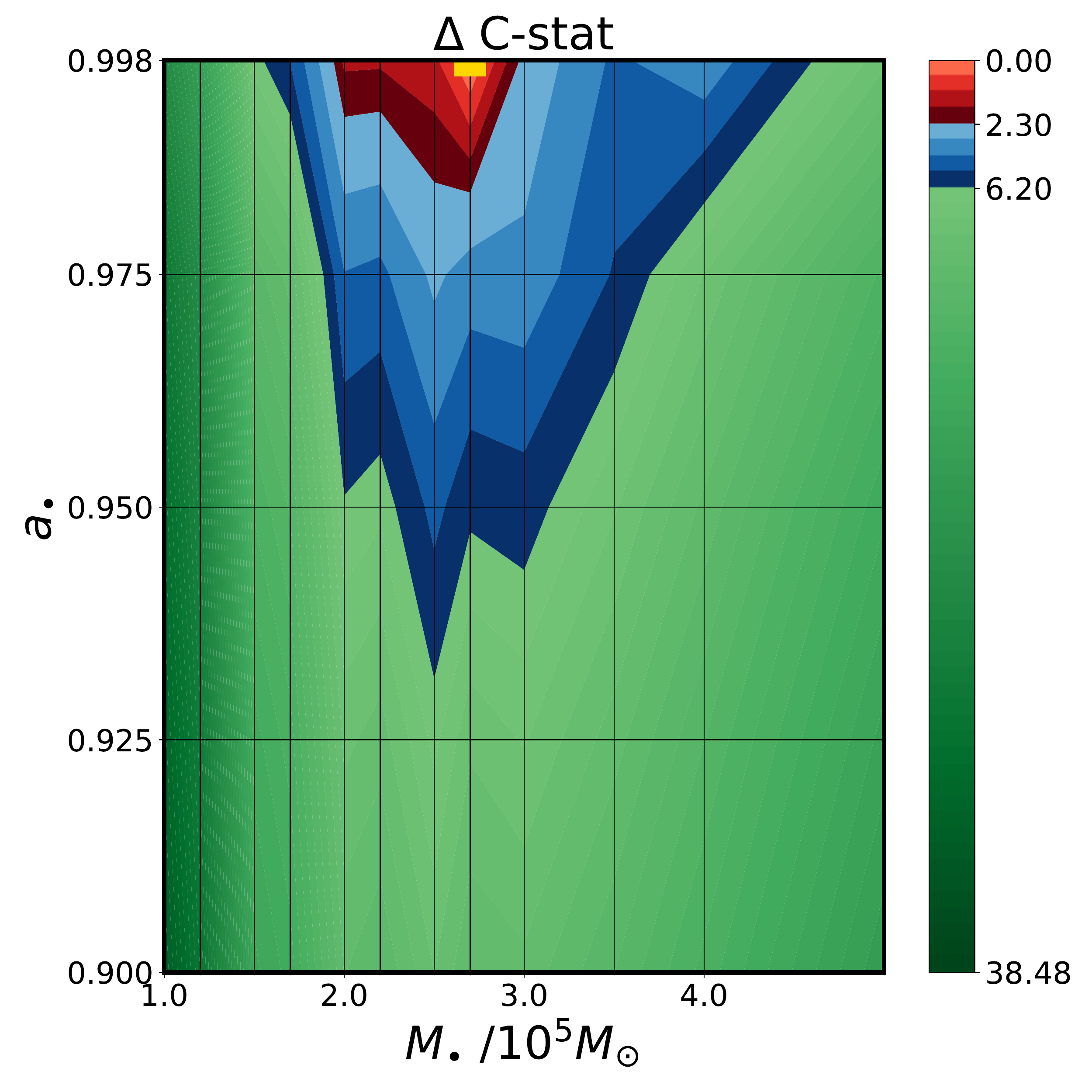}
    %\subfloat[\label{fig:hard-sp}]{\includegraphics[width=0.5\linewidth]{figures/contour-hard-sp.pdf}}\hfill
    \caption{Same as Fig.~\ref{fig:contour} but with $f_c$ parameter free--to--vary between 1 and 2.4 for late epoch spectra. The best--fit grid--point moves to a higher value of $M_{\bullet}$=$2.7\times10^{5}$ compared to Fig.~\ref{fig:contour} but the constrained 1$\sigma$ error range is not changed essentially for either $M_{\bullet}$ or $a_{\bullet}$, and the $\Delta$AIC=0 (C-stat/d.o.f.~$= 2023/2163$). Choices of $f_c$ for the late--epoch spectra do not influence the constraints on either $M_{\bullet}$ or $a_{\bullet}$ significantly.}
    \label{fig:contour-sp}
\end{figure*}

\begin{figure*}
    \centering
    \includegraphics[width=.5\linewidth]{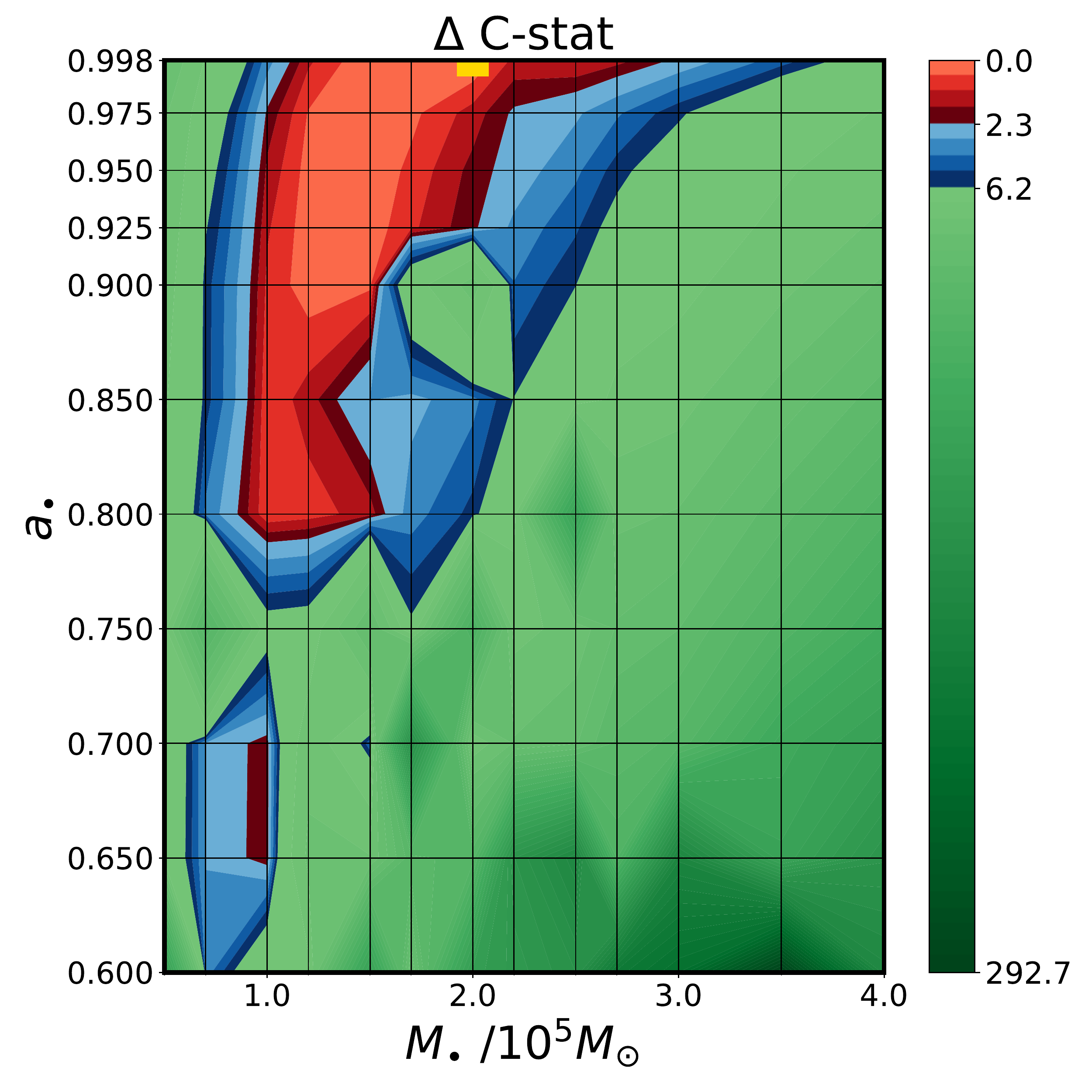}
    %\subfloat[\label{fig:hard-sp}]{\includegraphics[width=0.5\linewidth]{figures/contour-hard-sp.pdf}}\hfill
    \caption{Same as Fig.~\ref{fig:contour} but here the joint-fit uses only the data from late epochs (C9, X3, X4, X5, X6) and a fit--function of \texttt{TBabs*zTBabs*slimdisc}. Note the range in the y--axis are different from Fig.~\ref{fig:contour}. The best--fit C-stat/d.o.f.~$=707/748$. We find that the best--fit \{$M_{\bullet}$, $a_{\bullet}$\} values are the same as those derived considering both early-- and late epoch spectra, although the uncertainties on the best--fit values increase.}
    \label{fig:contour-late}
\end{figure*}

\begin{figure*}
    \centering
    \includegraphics[width=0.5\textwidth]{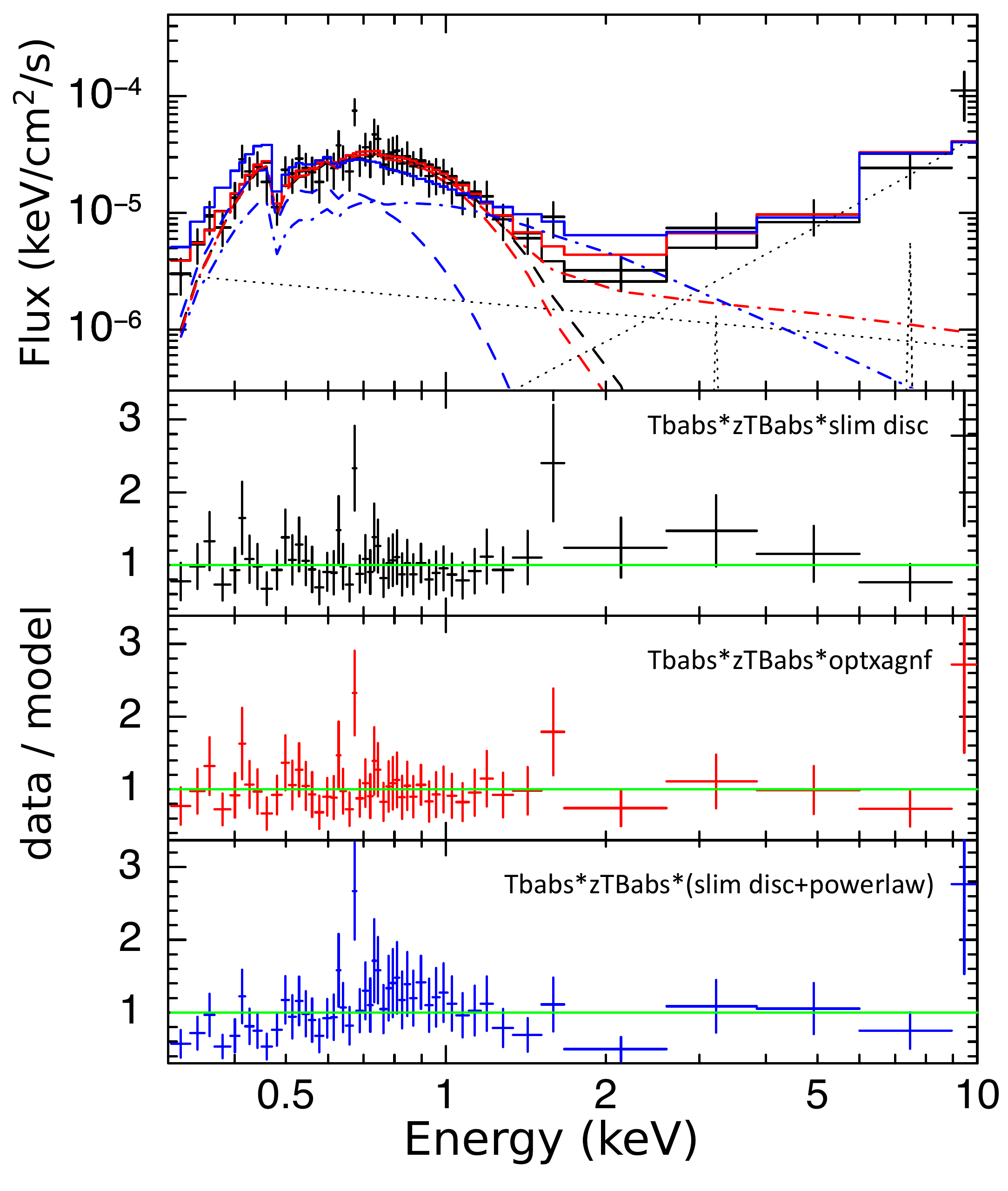}
    \caption{Comparison between our best--fit slim disc scenario (black) and \texttt{optxagnf} scenario (red) for the late--epoch spectra. For clarity we only show the X3 spectrum as an example. In \texttt{optxagnf} scenario, We set $N_{H,i}=0.26\times10^{22}$cm$^{-2}$, $M_{\bullet}=7.6\times10^{5}$~$M_{\odot}$, and $a_{\bullet}=0.998$, taken from \citealt{lin2022follow}.  \textit{Top panel}: the thermal disc emission (dashed lines) in \texttt{optxagnf} (red) and that in the slim disc model (black) are similar to each other. The red dot--dashed line represents the full \texttt{optxagnf} model (a thin disc$+$the non--thermal Comptonisation of a corona). We also include a scenario where the slim disc is forced to have $M_{\bullet}=8\times10^{5}$~$M_{\odot}$, and $a_{\bullet}=0.998$ (in blue color). In this case a powerlaw (blue dot--dashed line) is required to fit the data, resulting in C-stat/d.o.f.~$=761/748$. The same background models in all scenarios are represented by black dotted lines. The solid lines represent the total background$+$source model in each scenario. \textit{Rest panels}: The data$/$model ratio for each scenario.}
    \label{fig:opt-late}
\end{figure*}

\begin{figure*}
    \centering
    \includegraphics[width=.5\linewidth]{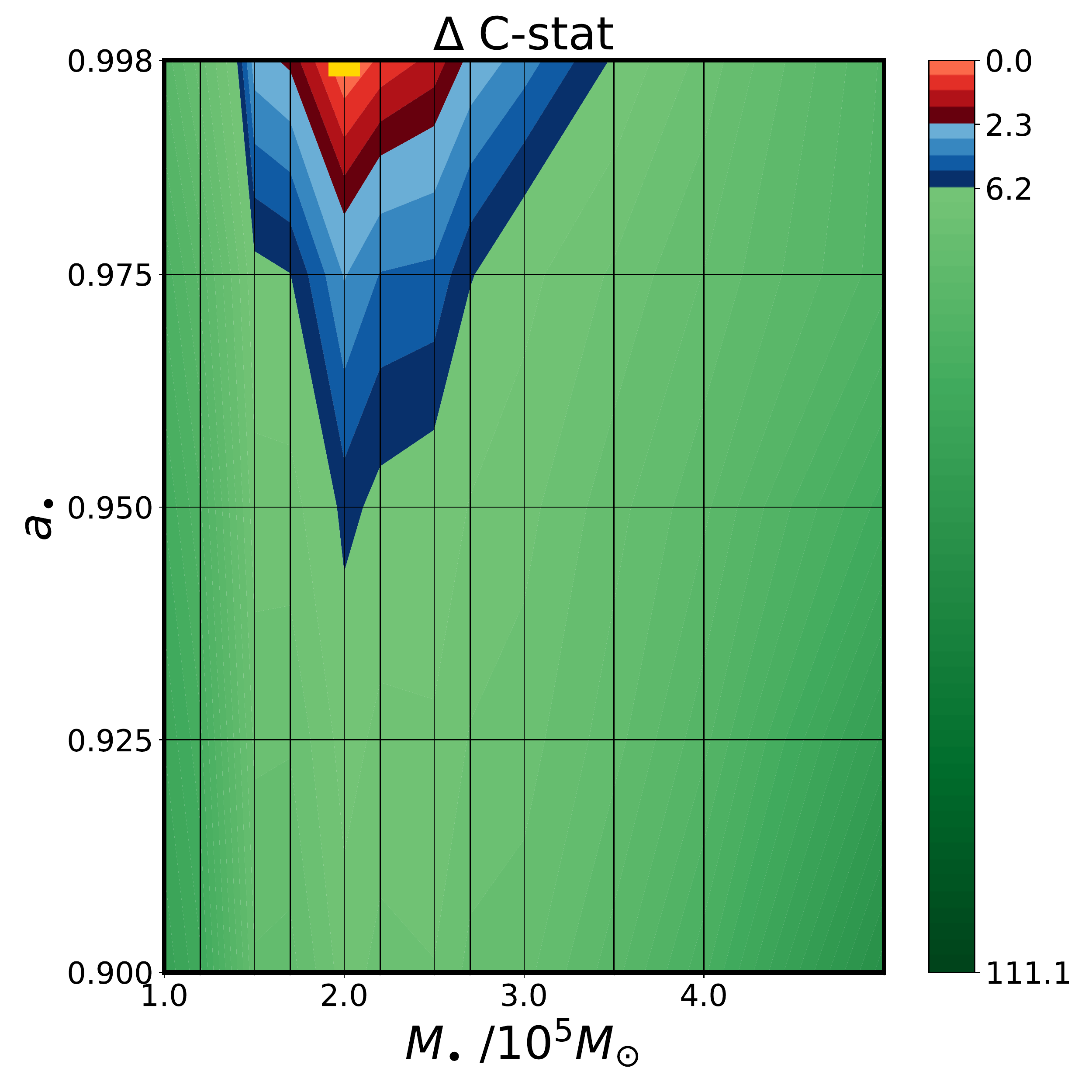}
    %\subfloat[\label{fig:hard-sp}]{\includegraphics[width=0.5\linewidth]{figures/contour-hard-sp.pdf}}\hfill
    \caption{Same as Fig.~\ref{fig:contour} but with the model \texttt{simpl} replacing the model \texttt{thcomp}, describing the up--scattered continuum by a power-law instead of assuming any specific electron energy distribution. The best--fit grid--point is the same as that in Fig.~\ref{fig:contour}. The constrained 1$\sigma$ error range is similar to that derived from the physically self-consistent thermal Comptonisation model \texttt{thcomp}. The $\Delta$AIC value is 4 (C-stat/d.o.f.~$=2020/2164$). We find the \{$M_{\bullet}$, $a_{\bullet}$\} constraints are not sensitive to whether or not the inverse--Comptonisation is done by electrons that have a thermal distribution.}
    \label{fig:contour-simpl}
\end{figure*}

%%%%%%%%%%%%%%%%%%%%%%%%%%%%%%%%%%%%%%%%%%%%%%%%%%

\newpage
% Don't change these lines
\bsp	% typesetting comment
\label{lastpage}
\end{document}